\documentclass[fleqn,usenatbib]{rasti}

\usepackage{newtxtext,newtxmath}

\usepackage[T1]{fontenc}

\DeclareRobustCommand{\VAN}[3]{#2}
\let\VANthebibliography\thebibliography
\def\thebibliography{\DeclareRobustCommand{\VAN}[3]{##3}\VANthebibliography}

\usepackage{graphicx}	
\usepackage{amsmath}	

\usepackage{subcaption}
\usepackage{textcomp, gensymb}
\usepackage{upgreek}
\usepackage{anyfontsize}

\usepackage{changepage}                 
\newlength{\offsetpage}
{\end{adjustwidth}}

\usepackage[usenames,dvipsnames]{xcolor} 

\title[Predicting RSO Populations]{Predicting RSO Populations Using a Neighbouring Orbits Technique}

\author[B. F. Cooke et al.]{
Benjamin F. Cooke,$^{1,2}$\thanks{E-mail: benjamin.cooke@warwick.ac.uk}
James A. Blake,$^{1,2}$
Paul Chote,$^{1,2}$
James McCormac$^{1,2}$
and
Don Pollacco$^{1,2}$
\\
$^{1}$Centre for Space Domain Awareness, University of Warwick, Gibbet Hill Road, Coventry CV4 7AL, UK\\
$^{2}$Department of Physics, University of Warwick, Gibbet Hill Road, Coventry CV4 7AL, UK
}

\date{Accepted XXX. Received YYY; in original form ZZZ}

\pubyear{2024}

\begin{document}
\label{firstpage}
\pagerange{\pageref{firstpage}--\pageref{lastpage}}
\maketitle

\begin{abstract}
The determination of the full population of Resident Space Objects (RSOs) in Low Earth Orbit (LEO) is a key issue in the field of space situational awareness that will only increase in importance in the coming years. We endeavour to describe a novel method of inferring the population of RSOs as a function of orbital height and inclination for a range of magnitudes. The method described uses observations of an orbit of known height and inclination to detect RSOs on neighbouring orbits. These neighbouring orbit targets move slowly relative to our tracked orbit, and are thus detectable down to faint magnitudes. We conduct simulations to show that, by observing multiple passes of a known orbit, we can infer the population of RSOs within a defined region of orbital parameter space. Observing a range of orbits from different orbital sites will allow for the inference of a population of LEO RSOs as a function of their orbital parameters and object magnitude.
\end{abstract}




\section{Introduction}
\label{sec:Introduction}


The number of Resident Space Objects (RSOs) in Earth orbit is increasing year on year \citep{MARK2019194,Blake2022,BERNHARD20231140}. There are databases which attempt to track the population of these objects (e.g. \href{https://www.space-track.org}{space-track.org}) but the fraction which are fully characterised falls off dramatically with size. In Low Earth Orbit (LEO) particularly, targets smaller than ${\sim}\,10$\,cm have low completeness \citep{2021AcAau.184...11P}. These targets are hard to track since most are not placed in orbit by design, but are debris associated with specific launches or satellites. Optical observations of these objects rely on reflected sunlight, thus their magnitude is proportional to their reflecting surface area and albedo. This means the smallest objects can be below the threshold of traditional observational methods.

The increase of objects in LEO is leading towards an increasing risk of in-orbit collisions \citep{2020AcAau.170...27P, 2023sndd.confE..67A}. Already, in recent years there have been multiple accidental collisions (as well as a few anti-satellite demonstrations), each of which has increased the budget of LEO debris by many thousands of pieces \citep{2010amos.confE..37S, 2011AdSpR..48..557P}. Due to their high relative velocities, LEO object collisions can produce catastrophic events \citep{2020AdSpR..65..351O}, both for the colliding objects, and by potentially increasing the density of debris to such an extent that entire sections of Earth orbit become unusable, known as the Kessler Syndrome \citep{1978JGR....83.2637K}. Avoiding future collisions requires a well populated database of RSOs, detailing their physical parameters and, importantly, their orbital elements \citep{2023amos.conf...77B}. A knowledge of the debris population will afford satellite operators the chance to perform orbital manoeuvres to avoid collisions \citep{2006JGCD...29.1140S, 2022AsDyn...6..121U}. Accurate RSO orbital elements will improve the effectiveness and efficiency of these manoeuvres \citep{2023AdSpR..72.4132D}.

Because of the small size and high orbital velocity of many LEO objects, they can be challenging to observe optically. The most straightforward way to observe a population of objects is to use a stationary telescope and detect targets as they cross the field of view \citep[e.g.][]{2019amos.confE..52C, 2021AdSpR..67..360B}. However, even for targets of sufficient size that the flux would be detectable, the relative velocity between the target and a stationary telescope may be sufficiently high that the flux is spread over so many pixels that each one falls below the detectability limit of the telescope. There have been multiple attempts to resolve these issues \citep{2023Senso..23.9668S}, including use of machine leaning for detection \citep[e.g.][]{2023amos.conf..205W}, observing from space \citep[e.g.][]{2022AdSpR..70.3271C}, and exposure stacking; both targeted \citep{2023AdSpR..72.2064Z} and blind \citep{COOKE2023907}. This paper describes a new approach, utilising neighbouring orbits. To avoid the issue of high relative velocities we employ a tracking telescope set-up, targeting a well defined orbit. We then attempt to recover RSOs on similar orbits, which will result in much reduced relative velocities. The method is thus capable of fainter (and thus smaller) target detection and is more focused on particular orbital regimes.

The neighbouring orbits detection technique as discussed herein is broadly simulation-based. We focus on a simulation-based approach to reduce the dependence on significant amounts of observational data which can be both expensive and limited. As this paper is a discussion of the technique and its applications, and not a full survey, this is reasonable. When discussing target detection in detail, we use example data from a specific telescope set-up to demonstrate the capabilities and limitations of the technique. The use of observational data and an expanded survey will be the basis of future work.

This paper begins with a description of neighbouring orbits and orbit generation (sections \ref{sec:Orbital parameters} and \ref{sec:TLE generation}), before detailing the simulation in full (sections \ref{sec:Offset TLE simulation} and \ref{sec:Effect of height, inclination and latitude}). It then continues into a discussion of the method used to detect targets on these neighbouring orbits (sections \ref{sec:Detection} and \ref{sec:Injection/recovery tests}) and how to convert detections into an inferred population (section \ref{sec:Detections to numbers}). Finally, we present our simulation results and conclusions, highlighting some future work (sections \ref{sec:Results} and \ref{sec:Discussion and conclusions}).

To simplify, sections \ref{sec:Orbital parameters} and \ref{sec:TLE generation} outline our assumptions and definitions. The simulation is described in section \ref{sec:Offset TLE simulation} and the detection process described in section \ref{sec:Detection}. Section \ref{sec:Detections to numbers} then extrapolates detections into results which are described in sections \ref{sec:Results} and \ref{sec:Discussion and conclusions}. Section \ref{sec:Effect of height, inclination and latitude} builds upon the base simulation, expanding the approach to different orbits and observing locations. Section \ref{sec:Injection/recovery tests} further details our detection method, justifying our methodology and providing avenues for adjusting thresholds.

\section{Orbital parameters}
\label{sec:Orbital parameters}

\begin{figure}
    \centering
    \begin{subfigure}{0.49\columnwidth}
        \centering
        \includegraphics[width=\columnwidth]{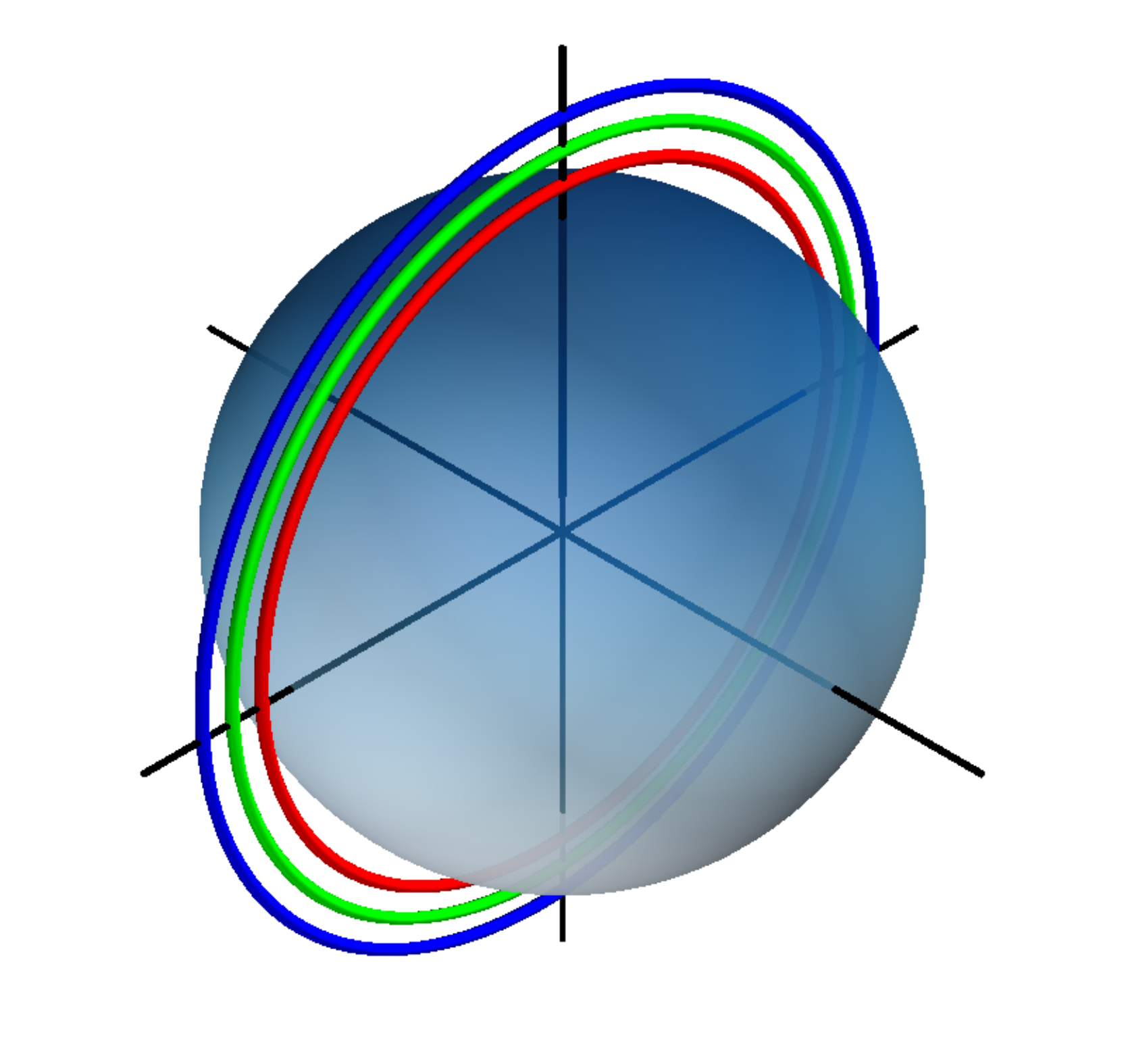}
        \caption{$h$ variation.}
        \label{fig:orbit_plot_height}
    \end{subfigure}
    \begin{subfigure}{0.49\columnwidth}
        \centering
        \includegraphics[width=\columnwidth]{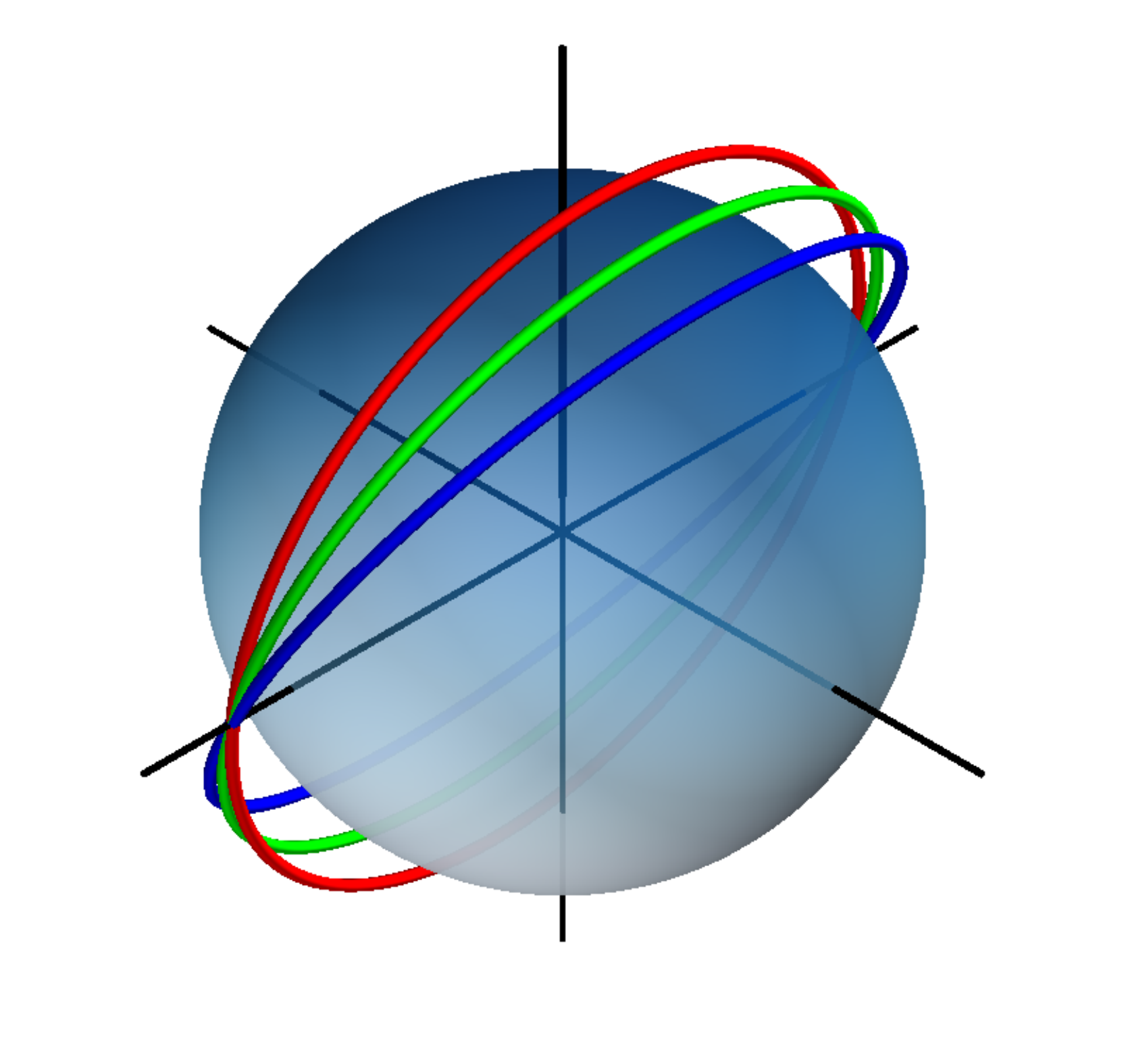}
        \caption{$i$ variation.}
        \label{fig:orbit_plot_inclination}
    \end{subfigure}
    \begin{subfigure}{0.49\columnwidth}
        \centering
        \includegraphics[width=\columnwidth]{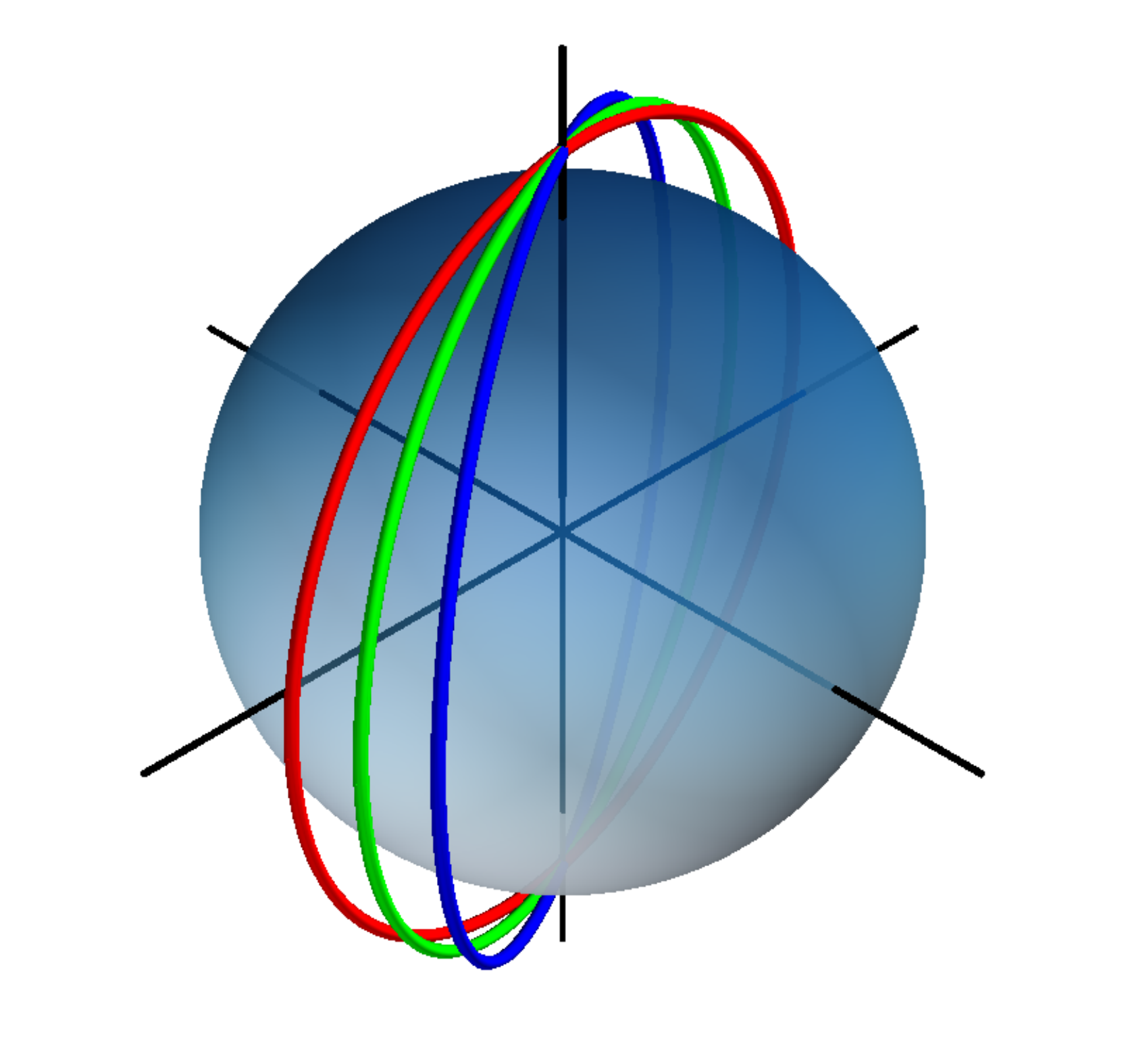}
        \caption{$\Omega$ variation.}
        \label{fig:orbit_plot_n}
    \end{subfigure}
    \begin{subfigure}{0.49\columnwidth}
        \centering
        \includegraphics[width=\columnwidth]{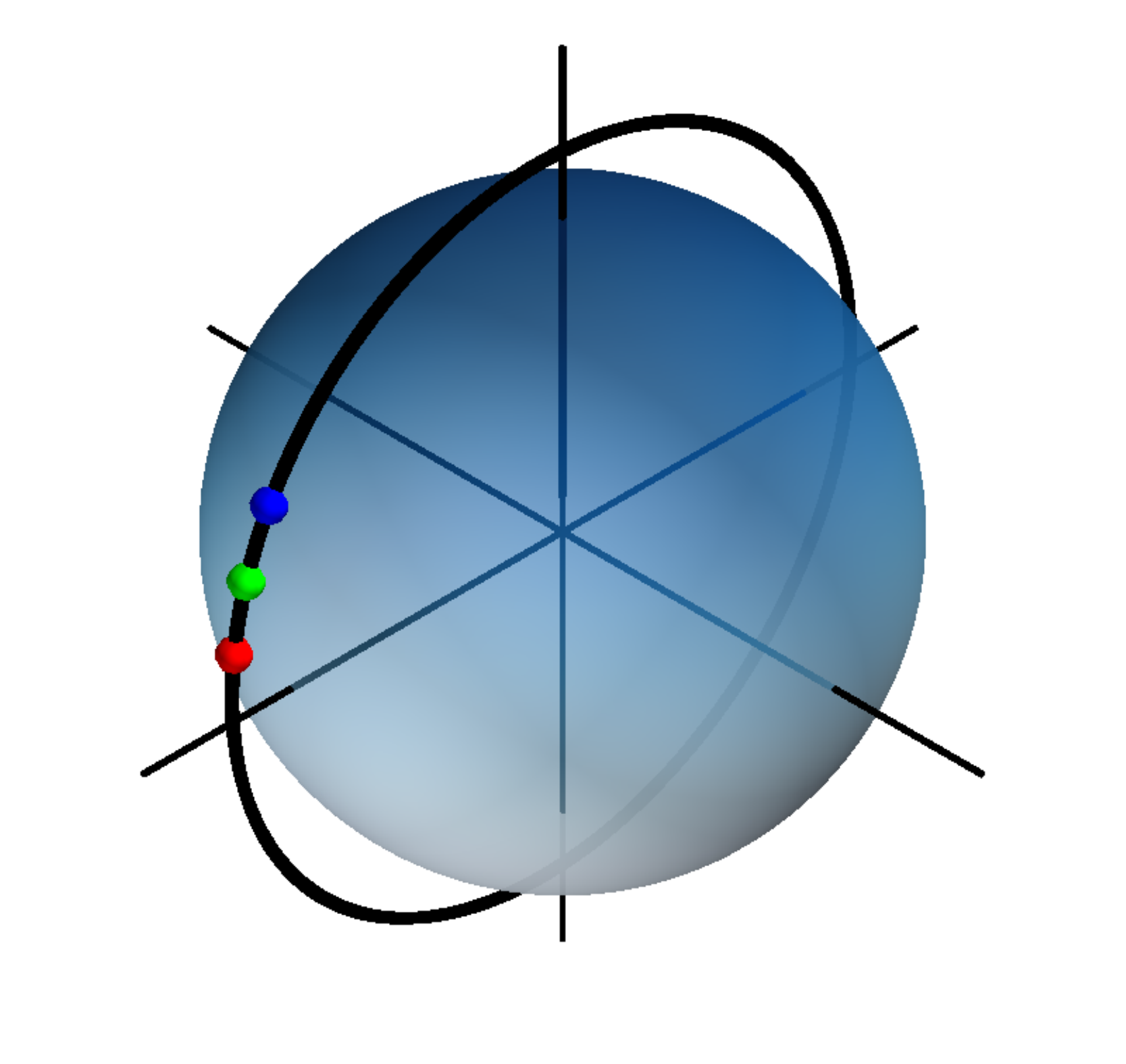}
        \caption{$\nu$ variation.}
        \label{fig:orbit_plot_m}
    \end{subfigure}
    \caption{Impact of orbital parameter variations. Coloured lines show orbits offset in the relevant parameter. For (d), coloured points show the difference in orbit start point, since the orbit itself is unchanged by the $\nu$ parameter. Lines progress from red to blue; $h$ ranges from $0.2-0.4R_\oplus$, $i$ ranges from $80-60\degree$, $\Omega$ ranges from $10-30\degree$, $\nu$ ranges from $10-30\degree$. Figures are generated using the \texttt{Mayavi} software package \citep{ramachandran2011mayavi}.}
    \label{fig:orbit_plot}
\end{figure}

To simplify our approach to this problem we make the assumptions that all orbits are circular (i.e. eccentricity, $e=0$). This means that we can define an orbit, and an object's position on it, using four key parameters. These four parameters are height $h$, inclination $i$, a modified right ascension of the ascending node $\Omega$, and a modified true anomaly $\nu$ (modifications are based on the chosen observing site and are discussed in section \ref{sec:TLE generation} and appendix \ref{sec:Orbital parameter derivations}). For computational ease, as well as simplifying the definition of other orbital parameters, we define a height $h$ and then set the apoapsis and periapsis equal to $h+R_\oplus\pm\epsilon$ where $\epsilon$ is a negligibly small value so as to simulate a circular orbit without forcing $e=0$. Since the orbit is, in practice, circular, $\nu$ does not actually affect the orbit and is instead used to assign a given object to a particular part of its orbit, allowing us to compare targets on identical orbits but with different epochs. Figure \ref{fig:orbit_plot} displays the effect of changing the parameters $h$, $i$, $\Omega$ and $\nu$ on a satellite's orbit. We define an orbit $O$ as a unique combination of the four orbital parameters which is written as $O = (h, i, \Omega, \nu)$ with $h$ given in units of kilometres and $i, \Omega, \nu$ in units of degrees.

\section{TLE generation}
\label{sec:TLE generation}


To use orbital parameters to predict the position of an RSO at any given time, we first generate a Two Line Element set (TLE). A TLE is simply a data format which encodes the specific parameters of an orbit. Specialised codes can then be used to predict the coordinates of an object on said orbit at any point in time. To generate the required TLEs we use the \texttt{faketle} tool from \texttt{SATTOOLS} \citep{sattools} which requires the four orbital parameters described above, as well as the epoch. We wish the generated orbit to have height $h$ and inclination $i$ and to pass directly overhead (i.e. across the zenith) of our observing site (with latitude=$\delta$ and longitude=$\alpha$) at a given epoch. We can use these parameters ($h$, $i$, $\alpha$, $\delta$, epoch) as well at the zenith crossing requirement, to determine the final two parameters $\Omega$ and $\nu$. $\Omega$ and $\nu$ are given by the following equations;

\begin{equation}
    \Omega = \phi_{RAAN} + \alpha - \phi_e
\end{equation}
\begin{equation}
    \nu = \cos^{-1}\left(\cos{\delta}\cos|{\alpha - \phi_e}|\right)
\end{equation}

where

\begin{equation}
    \phi_e = 2\tan^{-1}\left({\frac{\sqrt{A^2 + B^2 - C^2} - B}{A+C}}\right)
\end{equation}

and $A = \sin{\alpha}\cos{\delta}\cos{\theta}$, $B = \cos{\alpha}\cos{\delta}\cos{\theta}$, $C = \sin{\theta}\sin{\delta}$. For a full definition and derivation, see appendix \ref{sec:Orbital parameter derivations}.

\begin{figure}
    \centering
    \begin{subfigure}{\columnwidth}
        \centering
        \includegraphics[width=\columnwidth]{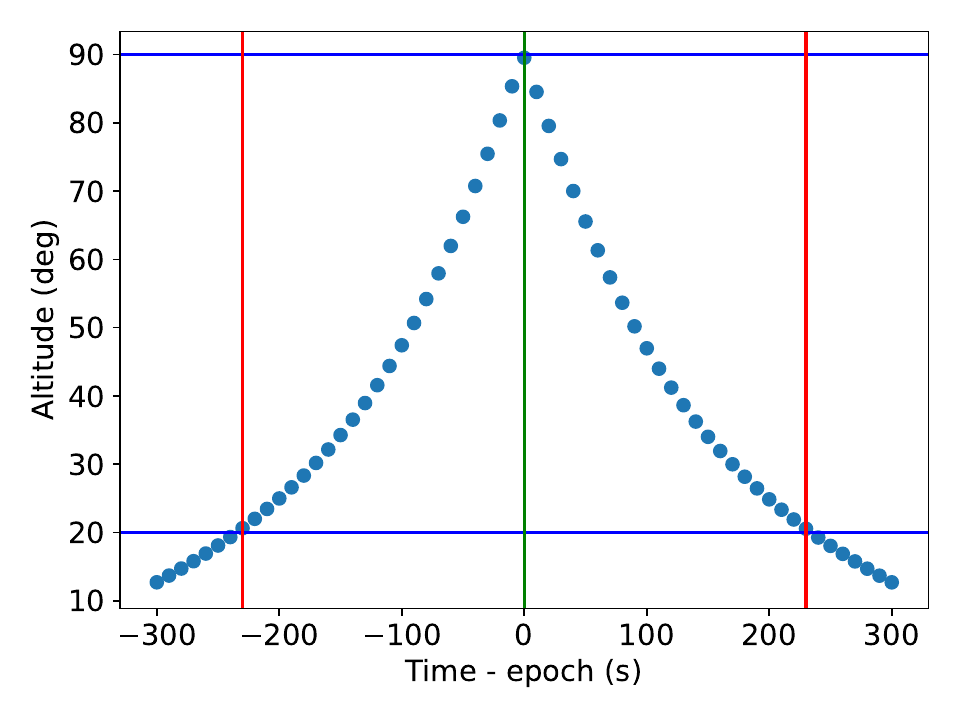}
        \caption{Target altitude as a function of time. $20\degree$ and $90\degree$ altitude lines are shown in blue, start and end of observable period are show in red with the defined epoch shown in green.}
        \label{fig:observing_constraints_alt}
    \end{subfigure}
    \begin{subfigure}{\columnwidth}
        \centering
        \includegraphics[width=\columnwidth]{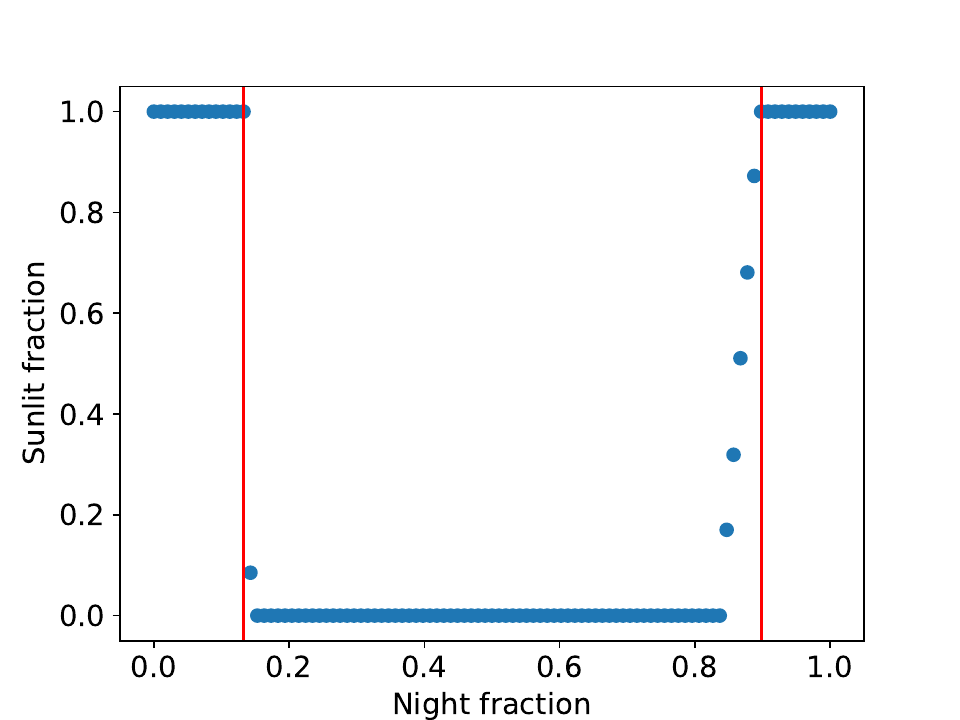}
        \caption{Fraction of orbit which is sunlit as a function of time, throughout a single night. The last, fully observable, evening pass and first, fully observable, morning pass are shown by the red lines.}
        \label{fig:observing_constraints_vis_time}
    \end{subfigure}
    \caption{Observing constraints for a target on an orbit with $h=850$\,km and $i=99\degree$ observed from a site on La Palma.}
    \label{fig:observing_constraints}
\end{figure}

The 5 orbital parameters are then used to generate the corresponding TLE which, in turn, is used to determine to coordinates of an object on said orbit at any point in time. The conversion of TLE to coordinates is done using \texttt{Skyfield} \citep{2019ascl.soft07024R}. The orbit is designed to cross the zenith of our observing site at the chosen epoch, but the amount of time for which it is observable before and after depends on both the observing site, and the specific orbital parameters. For a target to be observable it must satisfy two conditions. First, it must be at least $20\degree$ above the horizon of the observing site. This is due to a combination of instrumental and airmass restrictions. Second, it must be sunlit. For a target to be observable it must reflect sunlight which can then be seen by the telescope, therefore the vector from the sun to the target must not be intercepted by the Earth, while the sun must also be at least $6\degree$ below the horizon as seen from the observing site. Figure \ref{fig:observing_constraints} shows these criteria for an orbit with $h=850$\,km and $i=99\degree$, observed from our site on La Palma.

From Figure \ref{fig:observing_constraints_alt}, we see that our target is above the required altitude limits for 460\,s per pass, symmetrically around the chosen epoch. Figure \ref{fig:observing_constraints_vis_time} shows how the fraction of this period which is also sunlit changes throughout a night, beginning at fully observable after sunset, falling to fully unobservable during the middle of the night, and then rising again before sunrise. Successive passes are separated by 460\,s, the length of time for which a target is above $20\degree$ altitude. The number of fully observable passes per night is 25. Successive passes refer to successive observations of theoretical targets with the same orbital elements ($h$ and $i$) but different ascending nodes and true anomalies ($\Omega$ and $\nu$), not passes of the same target (which would be separated by $\sim90$\,min at LEO). Each pass tracks the path of a different theoretical target, which cross the meridian in 460\,s intervals ($\Omega$ and $\nu$ are adjusted to maintain a zenith-crossing object at an offset epoch).

\section{Offset TLE simulation}
\label{sec:Offset TLE simulation}


To determine whether an object in a neighbouring orbit is detectable, we need to generate a second TLE and a corresponding set of RA/Dec coordinates, at the same time stamps as used for the principal orbit. A neighbouring orbit refers to an orbit whose four orbital parameters are similar to those used in the principal orbit and are defined in terms of offsets from the principal orbit values. For each of the four orbital parameters we define a maximum offset value ($h_{\rm off}$, $i_{\rm off}$, $\Omega_{\rm off}$ and $\nu_{\rm off}$) and an offset step size ($h_{\rm step}$, $i_{\rm step}$, $\Omega_{\rm step}$ and $\nu_{\rm step}$). The maximum offset value is found using an iterative method and is such that increasing the offset of that orbital parameter any further (positively or negatively) will not result in any further detectable targets. It is dependent on the height and inclination of the orbit, and the latitude of the site, as discussed in section \ref{sec:Effect of height, inclination and latitude}. The offset steps are constant and chosen to be 2\,km in $h$ and $0.1\degree$ in $i$, $\Omega$ and $\nu$. We test both positive and negative offsets, thus the number of tested offsets for each parameter is given by $p_n = 2(p_{\rm off}/p_{\rm step}) + 1$ (where $p$ is a placeholder for any one of the four orbital parameters and the $+1$ term accounts for an offset of zero). We test all combinations of the four parameter offsets, resulting in $N_{\rm neighbours}$ neighbouring orbits being tested, where $N_{\rm neighbours} = h_ni_n\Omega_n\nu_n$.

\begin{figure}
    \centering
    \includegraphics[width=\columnwidth]{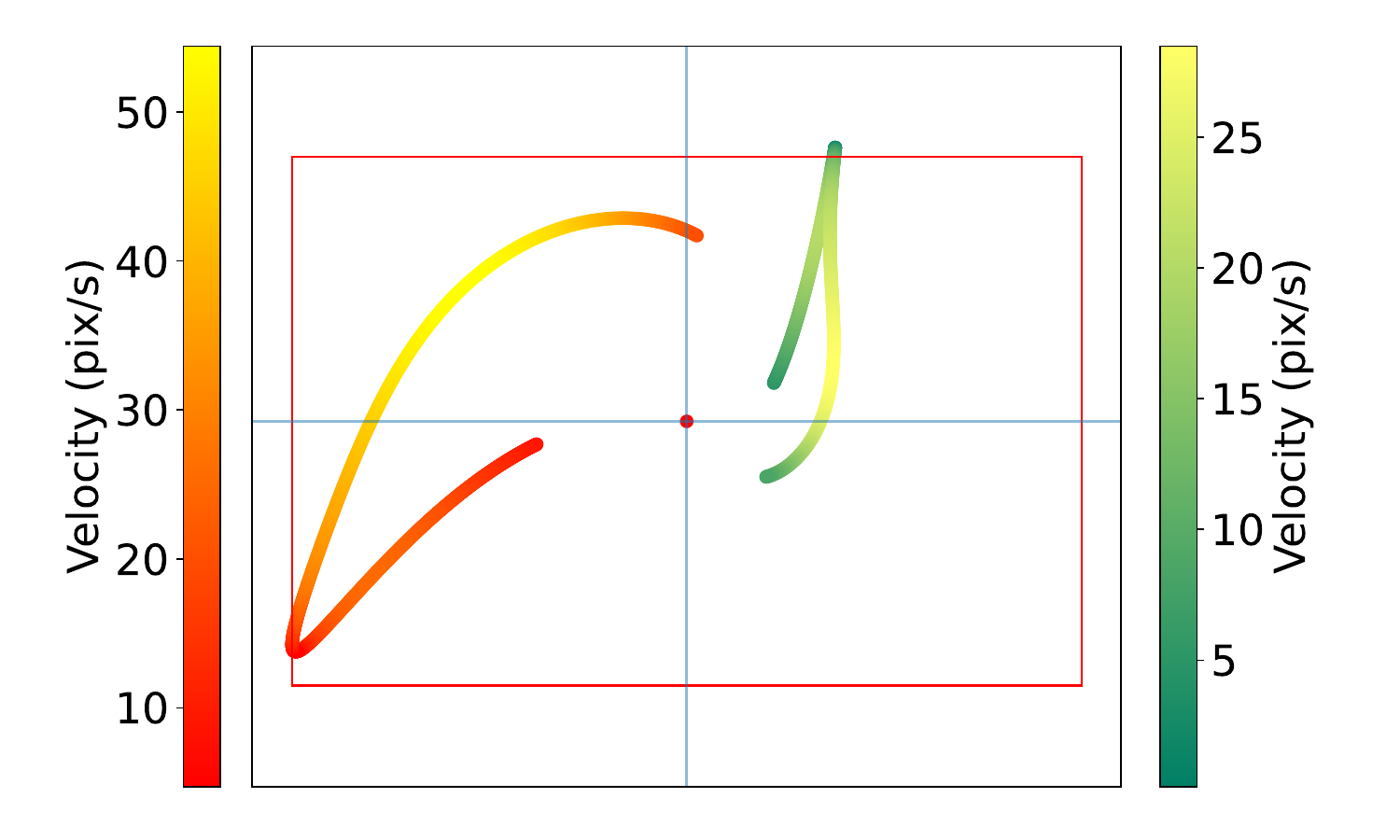}
    \caption{Location of targets on neighbouring orbits across the FoV of an instrument tracking the principal orbit for a full pass. The red point denotes the centre of the FoV (tracking $O_{\rm principal} = (850, 99, 29.8, 28.9)$) and the red box, its edges. The blue lines describe the central axes of the frame. The tracks of the two neighbouring targets are shown coloured by their velocities relative to the centre of the frame. Target 1, 
    $O_1 = O_{\rm principal} + (+2, +0.1, +0.1, -0.1)$, is on the right, coloured green/yellow and target 2, 
    $O_2 = O_{\rm principal} + (-2, +0.1, -0.1, +0.1)$ is on the left, coloured red/yellow.}
    \label{fig:neighboring_orbit}
\end{figure}

For each combination of offset orbital parameters we then produce a TLE as described in section \ref{sec:TLE generation} and generate RA/Dec coordinates for each of the time steps used for the principal orbit. We then convert these RA/Dec coordinates into frame positions, assuming that the frame is centred on the RA/Dec coordinates of the principal orbit. The equations are reproduced here, and described in full in \cite{radec_to_xy}, $(\alpha_1, \delta_1)$ are the RA/Dec coordinates of the principal orbit, and $(\alpha_2, \delta_2)$ are the RA/Dec coordinates of the neighbouring orbit (all coordinates are given in radians).


\begin{equation}
\label{eq:radec_to_x}
    x = x_{\rm sf}\left(\frac{\cos{\delta_2}\sin{(\alpha_2-\alpha_1)}}{\cos{\delta_1}\cos{\delta_2}\cos{(\alpha_2-\alpha_1)} + \sin{\delta_1}\sin{\delta_2}}\right) + x_c
\end{equation}
\begin{equation}
\label{eq:radec_to_y}
    y = y_{\rm sf}\left(\frac{\sin{\delta_1}\cos{\delta_2}\cos{(\alpha_2-\alpha_1)} - \cos{\delta_1}\sin{\delta_2}}{\cos{\delta_1}\cos{\delta_2}\cos{(\alpha_2-\alpha_1)} + \sin{\delta_1}\sin{\delta_2}}\right) + y_c
\end{equation}

where $x_{\rm sf}$ and $y_{\rm sf}$ are scale factors to convert radians to pixels given by

\begin{equation*}
\label{eq:xy_scale_factors}
    x_{\rm sf} = \frac{9600\times180}{2.63\pi} \\ y_{\rm sf} = \frac{6422\times180}{1.76\pi}
\end{equation*}

and $(x_c, y_c) = \left(\frac{9600}{2}, \frac{6422}{2}\right)$ are the coordinates of the centre of the frame (frame size used is for the CLASP instrument, specifications detailed in Table \ref{tab:instrument_specifications}). From the $x$/$y$ coordinates of the neighbouring orbit we can determine its relative separation from the principal tracked orbit and, assuming linear motion between successive time stamps, its relative velocity. Figure \ref{fig:neighboring_orbit} shows a plot of the relative motion of two targets on neighbouring orbits. 
The principal orbit is $O_{\rm principal} = (850, 99, 29.8, 28.9)$ and the two targets have orbits of $O_1 = O_{\rm principal} + (+2, +0.1, +0.1, -0.1)$ (right hand side, green/yellow) and $O_2 = O_{\rm principal} + (-2, +0.1, -0.1, +0.1)$ (left hand side, red/yellow). We see here how only relatively minor changes in the orbital parameters can drastically affect the resulting motion across the FoV.

For a target on a neighbouring orbit to be detectable it must satisfy two criteria. First, its $x$/$y$ location must place it within the bounds of a frame centred on a target on the principal orbit. That is, we require $0 \leq x < 9600$ and $0 \leq y < 6422$. Additionally, the relative velocity of the target must be below a certain threshold. Targets moving too fast, relative to the tracked target, will have their flux spread over too many pixels to be detectable using the detection method described in section \ref{sec:Detection}. The exact velocity threshold depends on the instrumental specifications and the detection goals and is discussed further in sections \ref{sec:Detection} and \ref{sec:Injection/recovery tests}, but the default value used here is equal to 10\,pix/s. We also require that these thresholds are met for a minimum number of consecutive frames. Based on the detection method described in section \ref{sec:Detection}, we require a target to be within the FoV and below the limiting relative velocity value for at least 20 consecutive frames. Additionally, we make the assumption that targets on neighbouring orbits are sunlit at the same time as a target on the principal orbit. This is a result of the relative speed of the TLE being significantly larger than the rotation speed of the Earth on the small FoV scale being considered. Simulations have confirmed that this is a reasonable assumption, which can significantly reduce computational runtime for testing orbital offset combinations.

\begin{figure}
    \centering
    \includegraphics[width=\columnwidth]{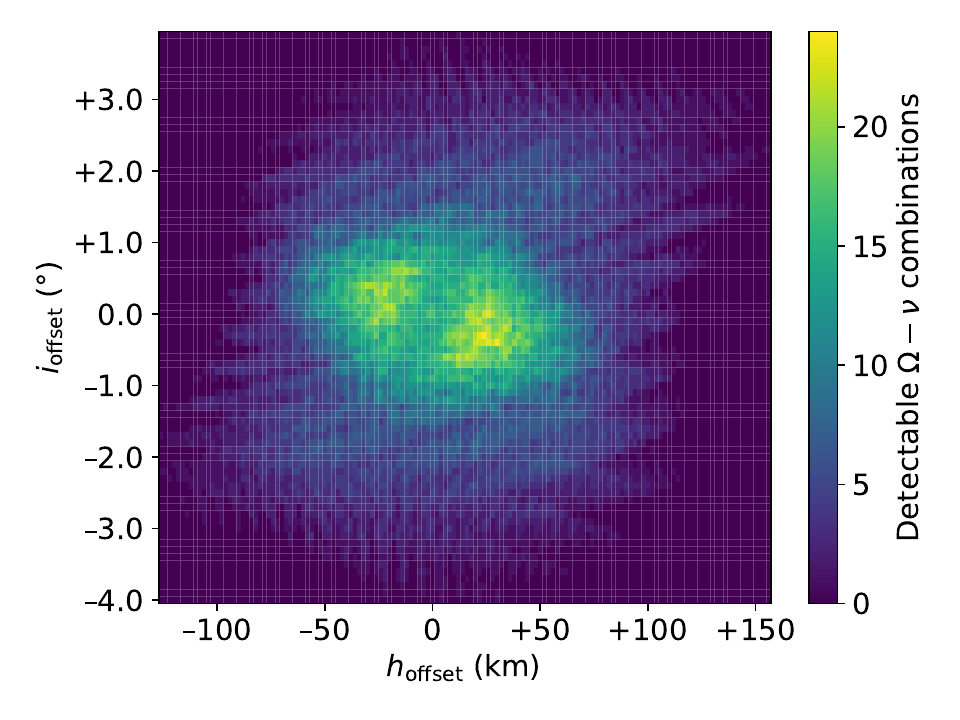}
    \caption{Target detectability as a function of $h_{\rm offset}$ and $i_{\rm offset}$ for 
    $O_{\rm principal} = (850, 99, 29.8, 28.9)$ observed from La Palma ($\delta=29\degree$) using CLASP. Colour designates relative probability, i.e. number of $\Omega-\nu$ combinations resulting in a detection for each $h-i$ pair. Values assume a limiting relative velocity of 10\,pix/s.}
    \label{fig:hi_detections_850_99_29}
\end{figure}

Each combination of offset parameters therefore produces a binary value of detectability. We don't account for targets which are detectable for a larger fraction of a pass than others, and simplify our results to detectable vs non-detectable offset parameter combinations. Figure \ref{fig:hi_detections_850_99_29} shows the resulting detectability map for a principal orbit with 
$O_{\rm principal} = (850, 99, 29.8, 28.9)$ observed from La Palma using CLASP. The results are presented in a 2D histogram in $h-i$ space with the colour representing the number of $\Omega-\nu$ combinations that produce a detectable target for each combination of $h-i$. For these parameters we find that detectable orbital parameters are as follows: 
$h = 850^{+156}_{-126}$\,km, $i = 99.0^{+3.9}_{-4.0}\degree$, $\Omega = 29.8^{+3.1}_{-3.2}\degree$, $\nu = 28.9^{+1.7}_{-1.4}\degree$. Note, however, that not every combination of offset parameters within these ranges will be detectable, just that these are the extreme values for each parameter independently.

The bimodality of target detectability seen in Figure \ref{fig:hi_detections_850_99_29} is due to the interplay between offsets in the four orbital elements. Near the very centre of the figure, the offsets in $h$ and $i$ are small, giving rise to an orbit very similar to the central one. Therefore, it takes only a small offset in $\Omega$ or $\nu$ to generate an orbit which is very similar in motion but sufficiently offset on the sky as to be outside the FoV and thus, undetectable. A slightly larger offset in either $h$ or $i$ gives a more significantly different orbit, creating a track which changes in position and velocity relative to the central one. Since the relative position and motion changes during the visible window, there are more combinations of $\Omega$ and $\nu$ offsets which produce a target orbit that is both sufficiently close and moving sufficiently slowly relative to the central one as to be detectable for at least a portion of this observing window. Increasing the $h$ and $i$ offsets further creates an orbit so different from the central one that no combinations of $\Omega$ and $\nu$ offsets can result in a target that is detectable.

The approximate symmetry is due to the fact that, on small scales, it matters not whether the compared orbit is offset positively or negatively to the central one. The relative difference between the central and offset orbit is broadly unchanged. The result is a broadly symmetrical bimodal distribution of detectability, peaking at low, but non-zero, $h$ and $i$ offsets, separated by a depressed region in the centre of the $h-i$ offset parameter space and surrounded by a region of low detectability at the extreme edges.

\section{Effect of height, inclination and latitude}
\label{sec:Effect of height, inclination and latitude}


The detectable ranges for $h_{\rm offset}$, $i_{\rm offset}$, $\Omega_{\rm offset}$ and $\nu_{\rm offset}$ are dependent on both the chosen principal values of $h$ and $i$, and the latitude of the observing site. The first impact is that the chosen values of $h$ and $i$ affect the amount of time for which a target is visible per orbit (we don't consider the effect of changing $\Omega$ or $\nu$ as they are not chosen directly, but calculated from $h$ and $i$). Larger values of $h$ mean higher orbits, and thus, slower moving targets, since the period of an orbit, $P \propto a^{3/2} = (R_\oplus + h)^{3/2}$. This means that a target remains above $20\degree$ for a greater amount of time per orbit. The number of fully observable passes per night remains roughly constant, since higher orbits mean that more time is required per pass, but also that a target remains sunlit for a greater amount of time after sunset and before sunrise. The combination of these two effects effectively cancel out. For example, observing an orbit with $h=750$\,km allows for 410\,s per pass and 26 passes per night. An orbit with $h=950$\,km allows 500\,s per pass and 25 passes per night. As described in section \ref{sec:TLE generation}, each pass tracks the path of different theoretical target (with the same $h$ and $i$ values), which cross the meridian in 410\,s (at $h=750$\,km) or 500\,s (at $h=950$\,km) intervals.

An increase in latitude of the observing site has no impact on the maximum observable time per pass (under the assumption that a pass crosses through the zenith), but it does affect how many fully visible passes are possible per night. 
The length of twilight depends on sun elevation and is longer closer to the poles, allowing for more passes per night. At a latitude of $29\degree$ we can observe 25 fully visible passes per night. At a latitude of $50\degree$ this increases to 33 passes per night and at a latitude of $75\degree$ we can observe 76 passes per night. 
It should be noted that these simulations were carried out during the northern hemisphere winter months at a solar declination angle of $-21\degree$ (close to mid-winter in the northern hemisphere, when the length of twilight is at its shortest). During the summer months twilight lasts longer and thus more observable passes per night would be expected.

\begin{figure}
    \centering
    \begin{subfigure}{\columnwidth}
        \centering
        \includegraphics[width=\columnwidth]{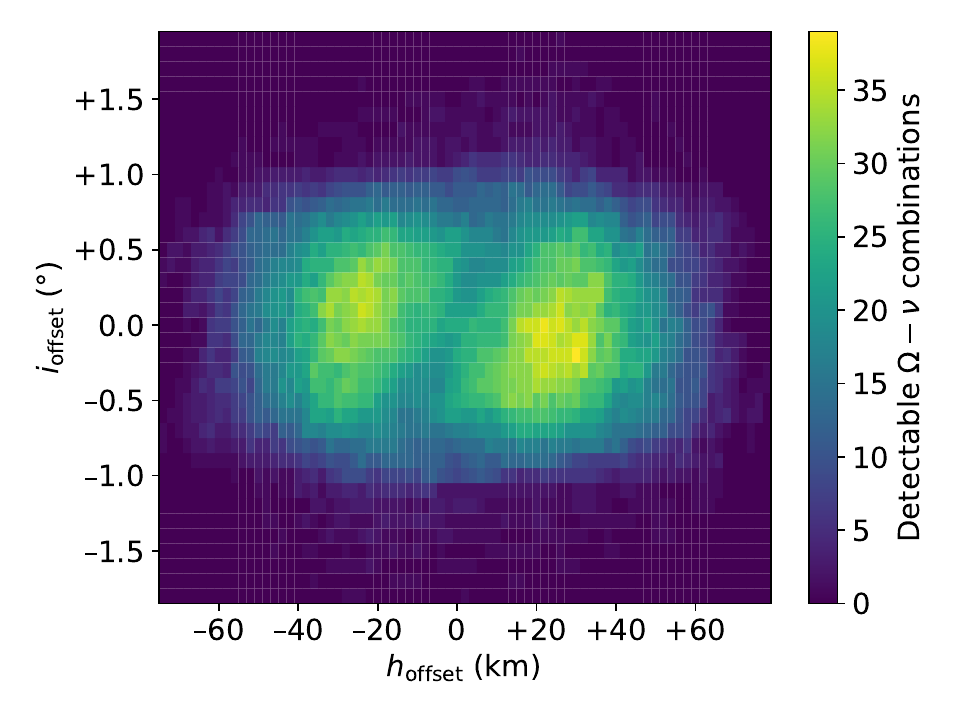}
        \caption{$\delta = 50\degree$}
        \label{fig:hi_detections_850_99_50}
    \end{subfigure}
    \begin{subfigure}{\columnwidth}
        \centering
        \includegraphics[width=\columnwidth]{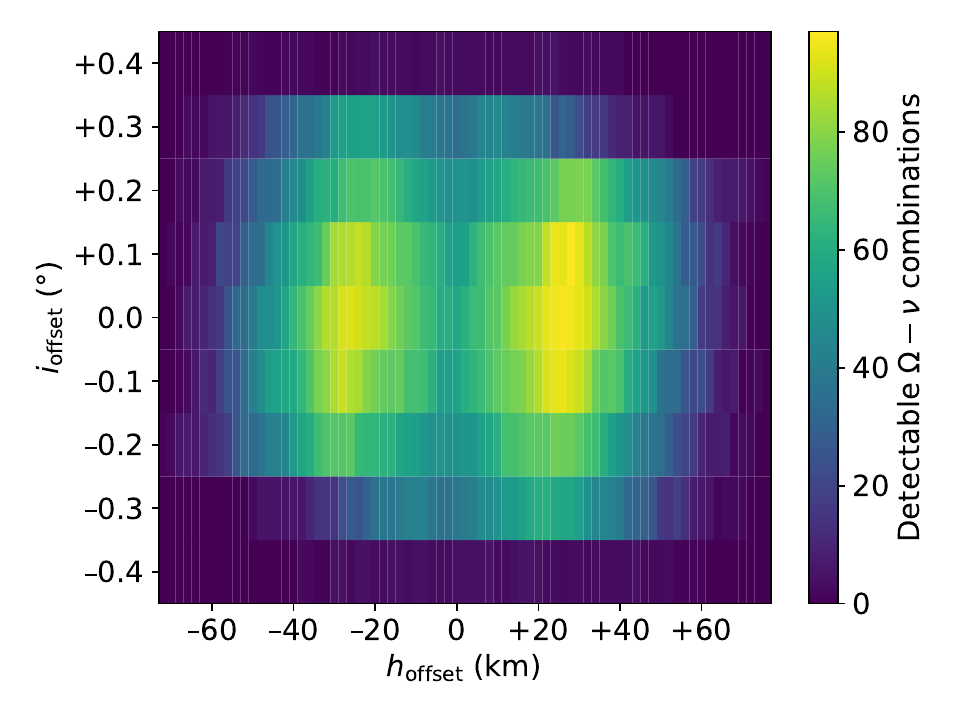}
        \caption{$\delta = 75\degree$}
        \label{fig:hi_detections_850_99_75}
    \end{subfigure}
    \caption{Target detectability as a function of $h_{\rm offset}$ and $i_{\rm offset}$ for 
    $O_{\rm principal} = (850, 99, 29.8, 28.9)$. Colour designates relative probability, i.e. number of $\Omega-\nu$ combinations resulting in a detection for each $h-i$ pair. Values assume a limiting relative velocity of 10\,pix/s. For ease of comparison, appendix \ref{sec:Target detectability comparison plots} reproduces figures \ref{fig:hi_detections_850_99_29}, \ref{fig:hi_detections_850_99_50} and \ref{fig:hi_detections_850_99_75} on the same axis scale.}
    \label{fig:hi_detections_850_99_50/75}
\end{figure}

Alongside the effects on the available observing time per pass, the chosen values of $h$, $i$ and $\delta$ can change the range of detectable orbital parameter offsets. Some of this is coupled to the change in observable time per pass. For example, consider a combination of orbital parameter offsets that produces a target that is only detectable for 10\,s, from $Z-230$ to $Z-220$ (where $Z$ is the time at which the tracked orbit reaches zenith). If the chosen $h$, $i$, $\delta$ values allow for only 410\,s of observation per pass (i.e. 205\,s before and after zenith) this offset orbit will not be considered detectable since the observable period begins 15\,s after the detectable period has ended. If, however, the chosen values allow for 500\,s of observation time per pass (i.e. 250\,s before and after zenith) the offset orbit will be considered detectable since the detectable portion occurs within the observable period. There are also visible time independent effects however. As discussed above, latitude has no effect on observable time per pass, but does affect the range of offset parameter combinations which produce detectable targets. Figure \ref{fig:hi_detections_850_99_50/75} shows the same sort of plot as displayed in Figure \ref{fig:hi_detections_850_99_29} but with different latitudes. The chosen 
orbit is $O_{\rm principal} = (850, 99, 29.8, 28.9)$ with $\delta = 50\degree$ (Fig. \ref{fig:hi_detections_850_99_50}), and $\delta = 75\degree$ (Fig. \ref{fig:hi_detections_850_99_75}) as opposed to $\delta = 29\degree$ used in Figure \ref{fig:hi_detections_850_99_29}. For ease of comparison, appendix \ref{sec:Target detectability comparison plots} reproduces figures \ref{fig:hi_detections_850_99_29}, \ref{fig:hi_detections_850_99_50} and \ref{fig:hi_detections_850_99_75} on the same axis scale.

As we can see from this figure, the chosen value of $\delta$ can significantly affect the resulting detectable ranges for $h_{\rm off}$, $i_{\rm off}$, $\Omega_{\rm off}$ and $\nu_{\rm off}$. 
We find the detectable ranges to be





\begin{equation*}
    O_{\delta=50\degree} = \left(850^{+78}_{-74},\, 99.0^{+1.9}_{-1.8},\, 29.8^{+3.0}_{-2.8},\, 28.9^{+0.7}_{-0.7}\right)
\end{equation*}

\begin{equation*}
    O_{\delta=75\degree} = \left(850^{+76}_{-72},\, 99.0^{+0.4}_{-0.4},\, 29.8^{+1.7}_{-1.7},\, 28.9^{+0.7}_{-0.7}\right).
\end{equation*}

\begin{figure}
    \centering
    \includegraphics[width=\columnwidth]{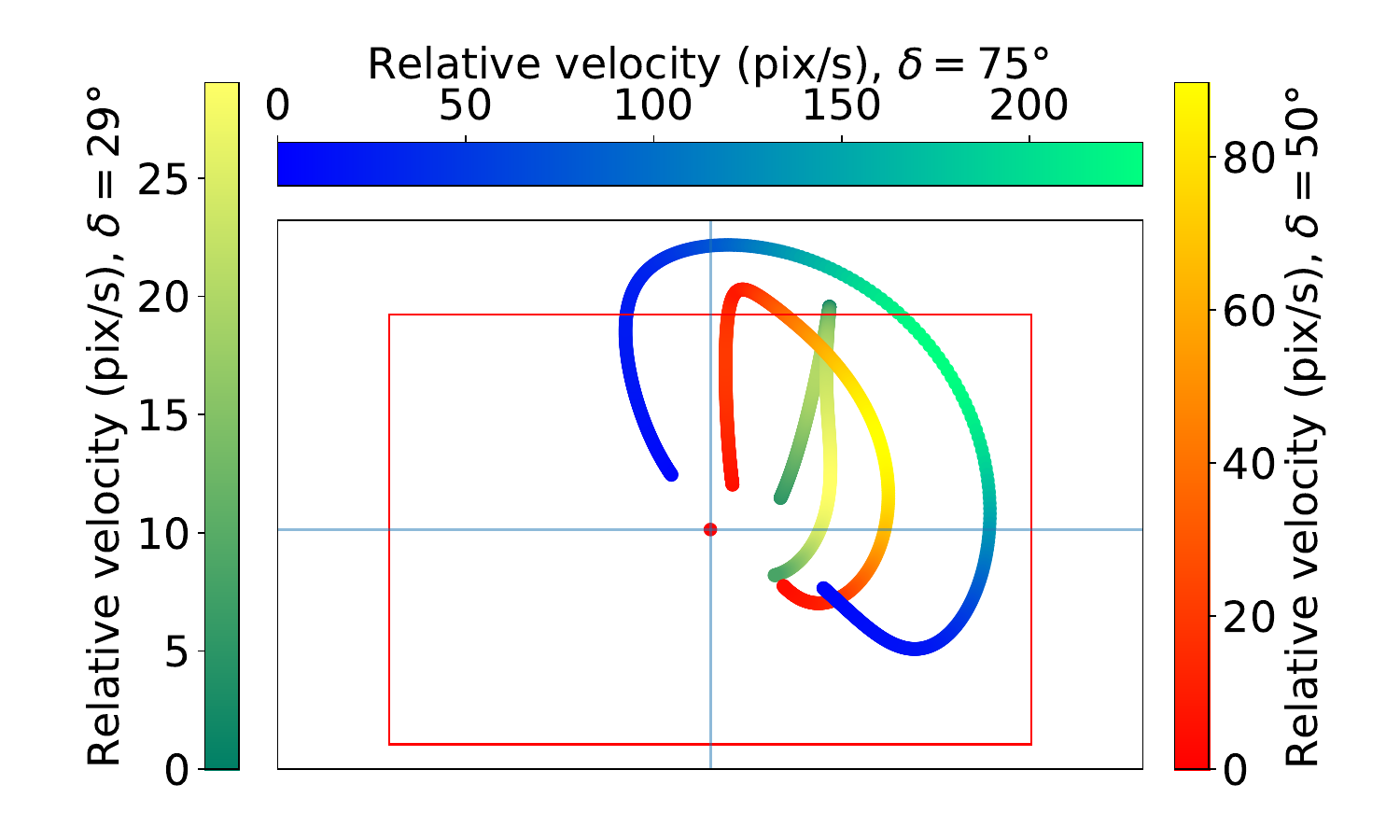}
    \caption{Relative motion of targets on neighbouring orbits as observed from different latitudes. General form is similar to Figure \ref{fig:neighboring_orbit}. The principal orbit 
    is $O_{\rm principal} = (850, 99, 29.8, 28.9)$ and offset orbit is $O = O_{\rm principal} + (+2, +0.1, +0.1, -0.1)$. The three tested latitudes are $\delta=29\degree$ (shown in green/yellow), $\delta=50\degree$ (shown in red/yellow) and $\delta=75\degree$ (shown in blue/green).}
    \label{fig:relative_motion_latitude}
\end{figure}

As $\delta$ increases, the detectable ranges of the four orbital parameters decreases. However, we see that although there are fewer allowed $h-i$ combinations, the density of detectable combinations actually increases. For example, we see that for $\delta=29\degree$ the $h-i$ combination with the highest density of $\Omega-\nu$ combinations has 24 valid $\Omega-\nu$ combinations. For $\delta=50\degree$, this increases to a maximum of 39 valid $\Omega-\nu$ combinations and for $\delta=75\degree$ we find a maximum of 97 valid $\Omega-\nu$ combinations. Figure \ref{fig:relative_motion_latitude} shows the relative motion of a target as observed from three different latitudes. 
The principal orbit is $O_{\rm principal} = (850, 99, 29.8, 28.9)$ and the neighbouring targets have $O = O_{\rm principal} + (+2, +0.1, +0.1, -0.1)$ The relative motions are calculated as observed from $\delta=29\degree$, $50\degree$ and $75\degree$.

The cause of the difference in detectable offset ranges is a combination of two major effects. The first is that, as $\delta$ increases, the relative offset between two orbits with different inclinations increases. Two orbits, differing in only inclination, will overlap where the orbit crosses the equator and have maximum separation furthest from the equator (see Fig. \ref{fig:orbit_plot_inclination} for a representation of this). Therefore, when observed from higher latitudes, an offset in inclination will have an increased effect of relative separation and motion between two targets, leading to a reduced inclination offset range for detectable targets. The second effect is that the relative offset between orbits with different $\Omega$ values decreases as the observing latitude increases. Orbits that differ only in $\Omega$ will have their maximum separation at the equator and minimum separation closest to the poles (see Fig. \ref{fig:orbit_plot_n} for a representation of this). Therefore, observing from a higher latitude site will result in a reduced effect from $\Omega$ offsets leading to a higher density of $\Omega-\nu$ combinations being detectable. These effects are complicated due to the fact that an offset in any single parameter effects the detectable offsets in every other parameter, but these factors are the driving forces behind the latitude dependent effects seen in figures \ref{fig:hi_detections_850_99_50/75} and \ref{fig:relative_motion_latitude}. The optimal choice of observing site depends on the exact goals of the survey and is discussed further in section \ref{sec:Discussion and conclusions}.

\section{Detection}
\label{sec:Detection}


Targets are identified in our observational data through the following method. Individual frames are first reduced using a combination of bias, dark and flat fields. Master bias and dark frames are obtained once, near the start of the observing period, and are compiled from an average of multiple individual frames. A master flat field frame is obtained every evening that observations occur, again, being an average of multiple individual frames. Observational data is gathered using the CLASP instrument on La Palma (see Table \ref{tab:instrument_specifications} for instrument specifications). 

To identify slow moving targets, i.e. those objects on similar orbits to our tracked orbit, and for the purposes of this project, we employ a frame stacking method. Frames are arranged sequentially and, after data reduction of each frame, we median stack rolling batches of $n$ frames where $n$ is an independent variable, with the default value being 10 (see section \ref{sec:Injection/recovery tests} for an in depth discussion of the optimal choice of variables). Each subsequent batch contains the last $n-1$ frames from the previous batch along with the next frame in the observation sequence. These stacked frames are then searched using the Python library for Source Extraction and Photometry \citep[\textsc{sep},][]{Barbary2016,1996A&AS..117..393B}, and we limit ourselves to smaller signals that have limited extension. Contamination from stellar signals are less of an issue, since they are mainly removed due to the stacking procedure and the limit of signal extension.

\begin{figure}
    \centering
    \includegraphics[width=\columnwidth]{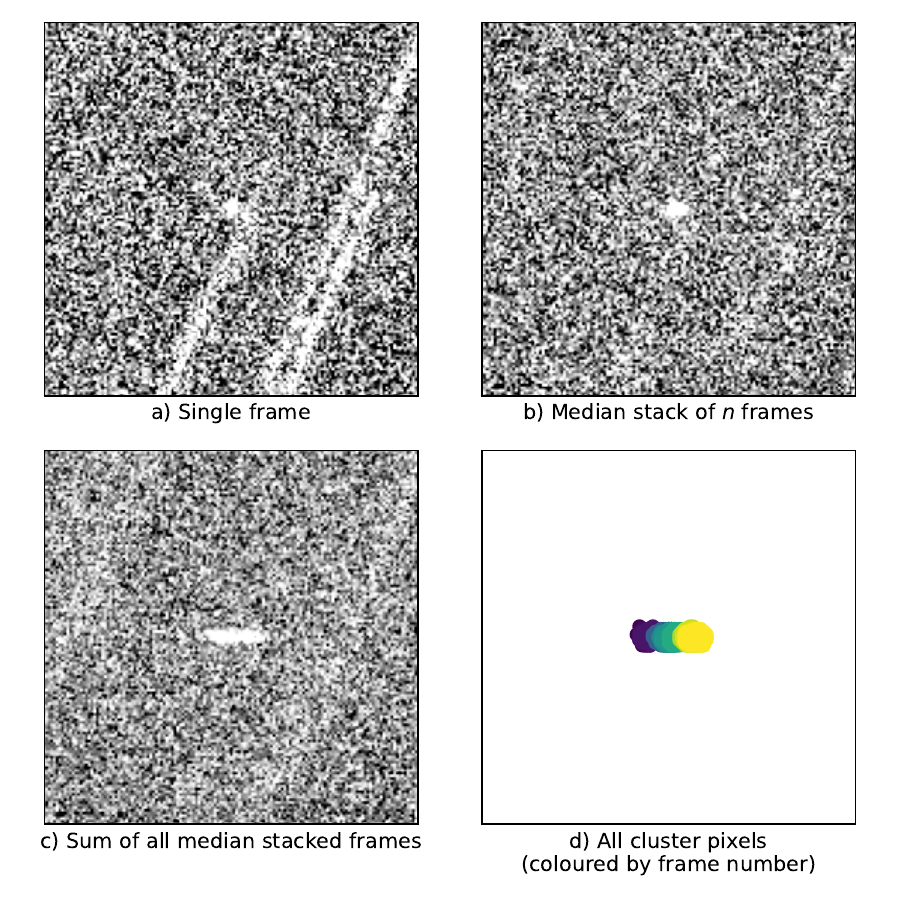}
    \caption{Detection of an injected signal. Panel a) single 0.5\,s exposure with injected signal. Panel b) median stack of $n=10$ frames. Panel c) sum of all stacked frames in which signal is detected. Panel d) Calculated cluster, with each cluster pixel coloured by the frame number in which it was identified.}
    \label{fig:paper_frames}
\end{figure}

For each detection, we record the detection parameters, as well as a list of the pixel coordinates of each pixel determined to be part of the detections. The results of SEP can be used to infer the magnitude and relative velocity of each detection. We then move into an additional step, required to limit false positives, and increase confidence in our detections. True detections are likely to appear in more than one frame and move across these frames in a predictable pattern, whereas false positives are more likely to appear in few, or single, frames and may appear to move in a more random pattern. To this end, we implement a bespoke clustering algorithm. For each detection, we define a physical and temporal buffer zone. We then search this buffer zone for other, similar, detections. If one is found, we assume it belongs to the same object as the original detection, and add it to the cluster of detections. The process is then repeated using this new detection as the origin of the `buffer' zone. This continues until no new detections are found, and we are left with a cluster of detections, similar in both physical and temporal space. Clusters which contain a suitable number of individual detections are determined to be real, and clusters which do not reach this threshold, are rejected as false positives. Figure \ref{fig:paper_frames} shows an example injected signal through each step in the process. Panel a) shows a single image with the injected signal, while panel b) shows a median stack of $n=10$ consecutive frames. Panel c) then shows the sum of all stacked frames in which the signal is detected and panel d) shows the calculated cluster, with each cluster pixel coloured by the frame number in which it was identified.

\section{Injection/recovery tests}
\label{sec:Injection/recovery tests}


It is important to know the theoretical faintest object that can be detected using the frame stacking method discussed above. The limiting magnitude is a function of the object's relative orbit (i.e. its relative speed with respect to the tracked orbit) and a range of observational and processing parameters including the number of threshold crossing pixels required for a detection ($n_{\rm pix}$), the number of frames per stack ($n$) and the exposure time of the frames ($t$). An additional approach that could affect detectability is binning pixels before attempting detection. This technique is not explored here but a discussion of pixel binning and its effects on the recovery of streaked targets is given in \cite{COOKE2023907}. Also important are a range of telescope-specific parameters such as the noise properties of individual frames ($\sigma$), as well as the Point Spread Function/Full-width Half Maximum (PSF/FWHM) and Zero Point (ZP). We use the CLASP telescope on La Palma for this work. The relevant Instrument specifications are given in Table \ref{tab:instrument_specifications}.

\begin{table}
\centering
\caption{Instrument specifications.}
\label{tab:instrument_specifications}
\begin{tabular}{c|c}
Parameter    & CLASP       \\ \hline
Diameter     & 36\,cm        \\
FoV          & $2.63\degree \times 1.76\degree$         \\
Frame size   & $9600 \times 6422$\,pixels     \\
Focal ratio & F/2.2       \\
Detector     & QHY600M sCMOS       \\
Pixel size   & 3.76\,$\upmu$m        \\
Gain         & 0.42\,e$^-$/ADU       \\
FWHM         & 3.6\,pixels       \\
Zero point   & 23.01\,mag
\end{tabular}
\end{table}

\begin{figure}
    \centering
    \includegraphics[width=\columnwidth]{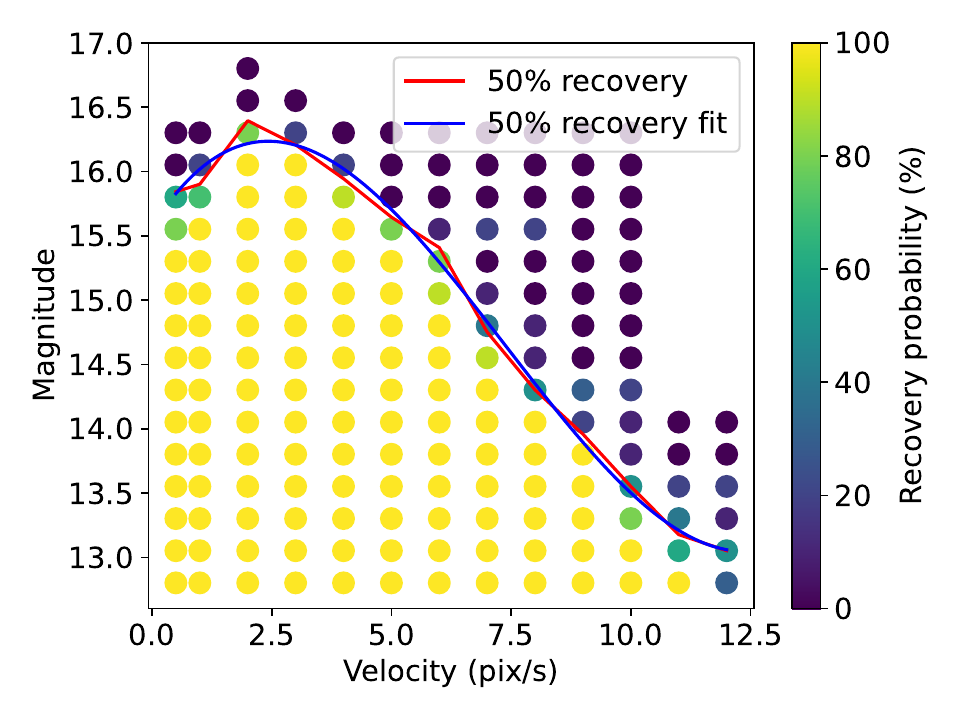}
    \caption{Recovery probability for injected signals with a range of magnitude and velocity values. The red line indicates a contour of 50\% recovery probability, assuming a linear interpolation between distinct points and the blue line shows a polynomial fit to the same data.}
    \label{fig:injection_recovery_prob}
\end{figure}

We have bench-marked this detection method through a range of injection/recovery tests. We simulate a range of signals, each corresponding to a target of a given magnitude, moving at a given relative velocity across the frame (streaks are simulated following the procedure laid out in \cite{COOKE2023907}). We then employ the above detection method to determine the reliability with which it can be recovered. Targets are injected multiple times, on different parts of the frame. The probability for recovery is simply the fraction of injections for a given magnitude/velocity combination that are recovered. Figure \ref{fig:injection_recovery_prob} shows the recovery probability for a tested set of injected signals spanning a range of magnitudes and velocities. The red line corresponds to a 50\% recovery fraction contour, assuming a linear interpolation between distinct points and the blue line shows a polynomial fit. These results employ $n = 10$, $n_{\rm pix} = 50$ and $t=0.5$s.

\begin{figure}
    \centering
    \begin{subfigure}{\columnwidth}
        \centering
        \includegraphics[width=\columnwidth]{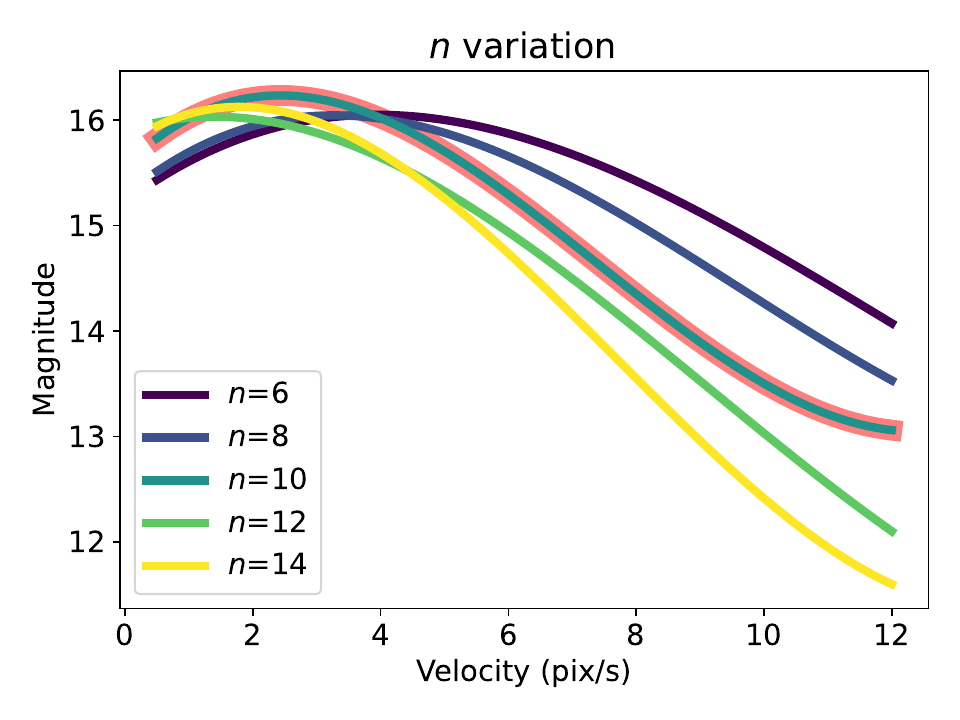}
        \caption{Number of frames.}
        \label{fig:mag_limits_n_frames}
    \end{subfigure}
    \begin{subfigure}{\columnwidth}
        \centering
        \includegraphics[width=\columnwidth]{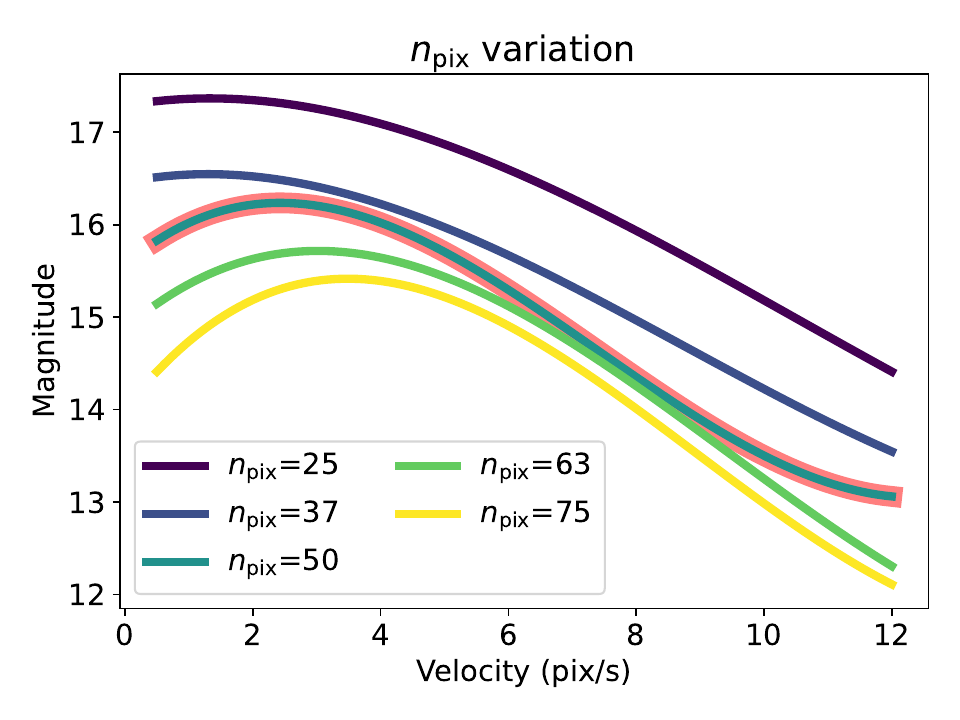}
        \caption{Number of pixels.}
        \label{fig:mag_limits_n_pix}
    \end{subfigure}
    \begin{subfigure}{\columnwidth}
        \centering
        \includegraphics[width=\columnwidth]{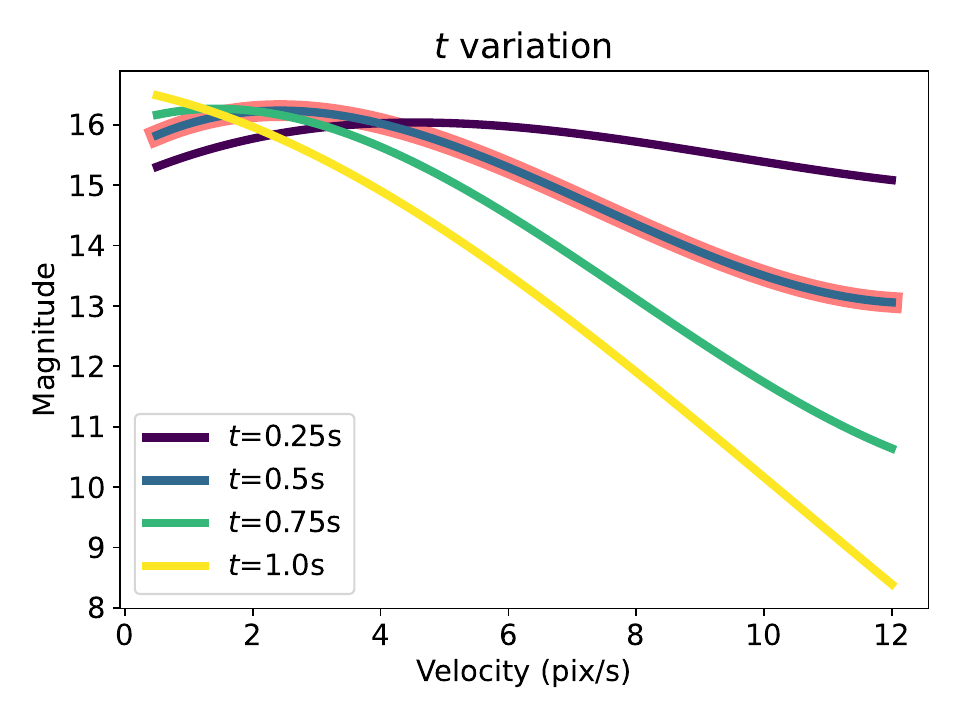}
        \caption{Exposure time.}
        \label{fig:mag_limits_exp_time}
    \end{subfigure}
    \caption{Limiting magnitude as a function of relative velocity of an object. Lines are polynomial fits to data. Sub-figures show how the relation is affected by different values of $n$, $n_{\rm pix}$ and $t$. The highlighted tracks corresponds to $n = 10,\ n_{\rm pix} = 50,\ t = 0.5$s.}
    \label{fig:mag_limits}
\end{figure}

From Figure \ref{fig:injection_recovery_prob} we see that the limiting magnitude for 50\% recoverability falls with increasing target velocity. This is to be expected, since the faster a target is moving, the more pixels its flux is spread over, and thus the fainter the signal it produces. Additionally, faster moving objects move across more pixels between subsequent exposures, making it less likely that stacking frames will be able to align the brightest parts of a signal, reducing the brightness of the signal in the stacked image. We also see that, somewhat counter-intuitively, the limiting magnitude also falls off for the slowest moving targets. This is due to the fact that the signals in subsequent images are so well aligned that the resulting signal in the stacked image is too small (in terms of number of pixels) to be identified by our detection method (to minimise the number of false positives we define a minimum number of pixels required for a detection, see Figure \ref{fig:mag_limits_n_pix}).

The best fit line for 50\% recovery, calculating magnitude $M$ as a function of velocity $v$, is a cubic, and has the form

\begin{equation}
\label{eq:M(v)}
    M = av^3 + bv^2 + cv + d
\end{equation}

with $a=0.006515$, $b=-0.1445$, $c=0.5864$ and $d=15.57$. Inverting the above equation to allow the determination of the maximum velocity for a given magnitude (and selecting for the correct cubic root) gives

\begin{equation}
\label{eq:v(M)}
    v = 2\sqrt{-\frac{p}{3}}\cos\left[{\frac{1}{3}\cos^{-1}\left({\frac{3q}{2p}\sqrt{\frac{-3}{p}}}\right)-\frac{2\pi}{3}}\right] - \frac{b}{3a}
\end{equation}

with $\displaystyle p=\frac{3ac-b^2}{3a^2}$ and $\displaystyle q=\frac{2b^3-9abc+27a^2(d-M)}{27a^3}$. For derivation of cubic roots, see \cite{cubic_roots}.

To determine how this limiting magnitude varies with the observational and processing parameters described above, we repeat the injection and recovery tests with a range of parameter values. Figure \ref{fig:mag_limits} shows how the limiting magnitude is affected by each of these variables in turn. For each sub-figure, two variables are held constant at their default values while the third is varied. Default values are defined as $n_{\rm pix} = 50,\ n = 10,\ t = 0.5$s. In each sub-figure the highlighted line (thicker line, outlined in pink) shows the limiting magnitude for the default values.

From Figure \ref{fig:mag_limits_n_frames}, we see that lowering the value of $n$ leads to the detection of fainter targets at the high velocity end of the distribution. This is due to the fact that targets need to remain on the same pixels for fewer frames, thus they can be correspondingly fainter and still be detected. However, at the slower velocity end, the detection limit actually decreases. This is due to the fact that the flux from these targets is now spread over so few pixels, that their extended size on a frame is below the threshold size for detection. Based on simulations (see Section \ref{sec:Offset TLE simulation}), we see that the relative velocity of an object changes during the observing window. Therefore, targets which are, at one point, moving too slowly to be detected, are likely to increase their relative speed during observations, moving them into a region of parameter space with a higher recovery probability, resulting in a successful recovery. Additionally, lowering the number of frames leads to a less successful removal of background stellar signals and thus an increase in the number of false positives. Increasing the value of $n$ has the opposite effect, with a brighter detection limit for faster moving targets and a slightly fainter limit for the slowest moving targets. Using $n=10$ gives the best trade off between magnitude limit and false positives.

From Figure \ref{fig:mag_limits_n_pix}, we see that lowering the value of $n_{\rm pix}$ leads to a fainter magnitude limit at all velocities. This is as expected, since allowing smaller signals to be picked up means that fainter targets, which have a smaller size on the frame, can be detected. The consequence is that reducing this threshold leads to a drastic increase in the number of false positive detections. Conversely, increasing the value of $n_{\rm pix}$ means that targets must be brighter to be successfully detected, but reduces the number of false positives. We find that $n_{\rm pix}=50$ gives a good compromise between increased detectability and a reduced number of false positives.

From Figure \ref{fig:mag_limits_exp_time}, we see that lowering the value of $t$ results in improved detections for the faster moving targets. 
This is due to shorter exposures meaning that targets move across fewer pixels in individual frames (trails are shorter). This increases the chance that they remain on similar parts of the image in successive frames, thus they are more likely to be detectable in stacked images. At the slower end of the velocity range the detectability falls off with shorter exposures, due to the reduced size of the signal on an image, moving it below our detection threshold. Reducing the exposure time also leads to a higher data rate. In a realistic scenario, data storage and processing time are important considerations, both of which increase as exposure time decreases. For longer exposure times, the data requirements are reduced but the detectability suffers due to the inverse of the effects mentioned above. Using $t=0.5$s gives a compromise between these considerations.

\section{Detections to numbers}
\label{sec:Detections to numbers}


To convert detections into a prediction for the number of objects within a certain region of orbital parameter space, we need a reasonable underlying distribution. One approach is to use a uniform distribution in the four orbital parameters described above, $h$, $i$, $\Omega$ and $\nu$. This however, is not realistic for all parameters, since some orbits are more favoured than others, for example, polar orbits (with $i\sim 90\degree$) are significantly more heavily populated than others. Additionally, depending on the purpose of particular space missions, some orbital heights are likewise more heavily utilised. To determine a reasonable underlying distribution, we utilised the ESA Meteoroid and Space Debris Terrestrial Environment Reference model v8.0.3 \citep[MASTER,][]{doi:10.2514/6.2016-5658}. MASTER uses a combination of confirmed objects with TLEs and simulations to produce a theoretical population of orbiting debris (MASTER also includes a population of meteoroids). For $\Omega$ and $\nu$ we assume a uniform distribution but for $h$ and $i$ we use the MASTER database to infer the underlying distribution of satellites and debris in Earth orbit. We use the MASTER population for all objects larger than 1cm in size, which is a combination of known objects and simulation. We assume that the underlying distribution of MASTER objects is a reasonable approximation to the true underlying distribution, even if the actual number of objects is less certain. Utilising the MASTER distribution will allow us to make our own inferences about the  number of targets in a particular region of parameter space.

\begin{figure}
    \centering
    \includegraphics[width=\columnwidth]{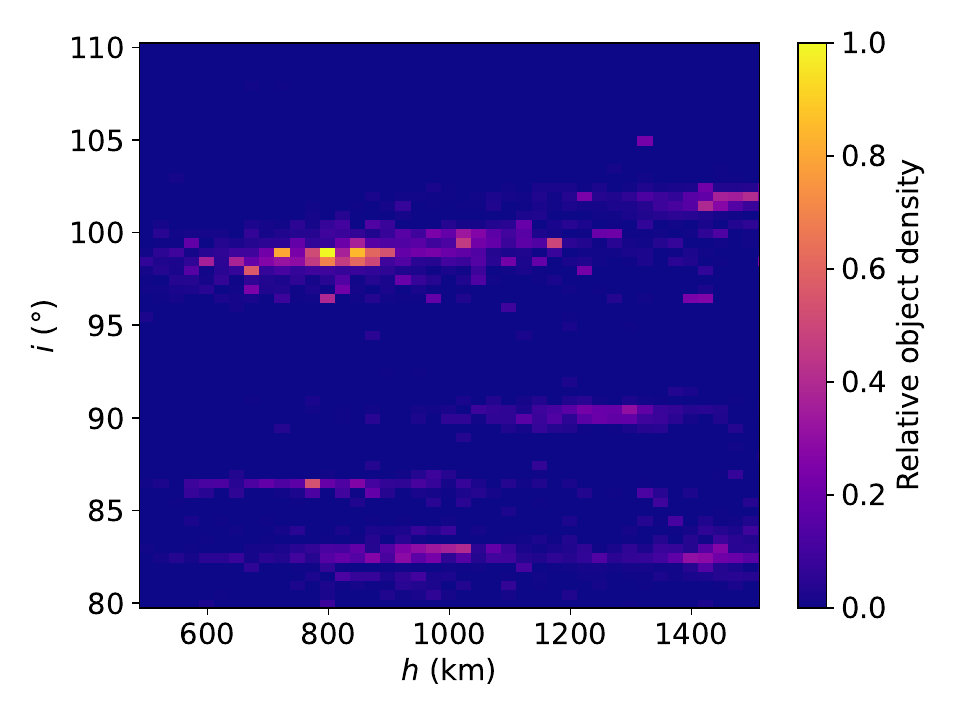}
    \caption{Relative number of objects as function of orbital height and inclination (full height and inclination ranges are truncated for visibility) from the MASTER population. Only objects $\geq1$\,cm in size are shown.}
    \label{fig:MASTER_density}
\end{figure}

On large scales we find significant structure in the distribution of $h$ and $i$ of RSOs. Figure \ref{fig:MASTER_density} shows a portion of a 2D histogram of the MASTER population $h-i$ parameter space, coloured by the relative number of objects in each bin.

In this figure we see that some $h-i$ bins are much more heavily populated than others. On smaller scales however, the distribution is much less structured. Within each bin shown in Figure \ref{fig:MASTER_density} the distribution of specific objects is much more random. Therefore, we assume that, though the number of objects per bin can be inferred from the MASTER population, the distribution of objects within each bin is uniform. The bin sizes we use as a separation between these two regimes are 25\,km in $h$ and 0.5$\degree$ in $i$.

To use this distribution to convert detections to targets we proceed in the following way. We assume $N$ objects total, within the tested region of $h-i$ parameter space. The extent of the $h-i$ region about which inferences can be made is a result of the minimum and maximum detectable values of $h_{\rm offset}$ and $i_{\rm offset}$ found by the simulations discussed in section \ref{sec:Offset TLE simulation}.

This region of parameter space is then split into $A$ offset parameter bins, each corresponding to a different tested combination of $h_{\rm offset}$ and $i_{\rm offset}$. For each of these offset parameter bins, we determine which MASTER population bin it falls into (the MASTER population bins are shown in Figure \ref{fig:MASTER_density}), noting that since the offset bins are smaller there will be $b = \frac{25}{2}\times\frac{0.5}{0.1} = 62.5$ offset bins within each MASTER population bin.

The fraction of the $N$ total objects that fall into each offset bin is given by $M_a/bM$ where $M_a$ is the number of MASTER population objects in the MASTER bin that contains the current offset bin $a$, and $M$ is the total number of MASTER objects in all $A$ offset bins within the tested $h-i$ parameter space. Thus each offset bin contains $N_a = NM_a/bM$ objects.

Within an individual parameter offset bin, the probability of any individual object being detected is proportional to the number of $\Omega-\nu$ combinations that result in a detectable target for the relevant $h-i$ pair. In fact, the exact probability is equal to the fraction of all possible $\Omega-\nu$ combinations that give a detectable result, i.e. $P_a = C_{a,\rm detectable}/C_{\rm total}$ where $C_{a,\rm detectable}$ is the number of detectable $\Omega-\nu$ combinations for $h-i$ bin $a$ and  $C_{\rm total} = (360/0.1)^2$, is the total number of $\Omega-\nu$ combinations, since the $\Omega$ and $\nu$ offset bins are 0.1$\degree$ wide and can have any value from $0\degree \leq \Omega,\nu < 360\degree$.

The expected number of detections within a single offset bin, $d_a$, is thus equal to the number of objects in the relevant bin, multiplied by their probability for detection, $d_a = N_aP_a$. The total number of detections $D$, is thus the sum of $d_a$ over all $A$ bins, $D = \sum{N_aP_a}$. Expanding the terms and solving for $N$, thus gives a prediction of the number of objects $N$ as a function of the number of detections $D$;

\begin{equation}
\label{eq:D_to_N_one_pass}
    N(D) = D\frac{bMC_{\rm total}}{\displaystyle \sum_{a=1}^{a=A} M_aC_{a,\rm detectable}}
\end{equation}

For a single pass the above equation holds true, i.e. tracking a TLE through one pass (as described in sections \ref{sec:TLE generation} and \ref{sec:Offset TLE simulation}) and making $D$ detections will result in a predicted $N$ objects in the relevant region of $h-i$ parameter space. For multiple passes however, the average number of detections per orbit should instead be used. Since the above detectable offset combinations refer to a detection during a single pass, detections on subsequent passes give further information about the underlying population. Therefore, across $p$ passes,

\begin{equation}
\label{eq:D_to_N_multiple_passes}
    N(D) = \frac{D}{p}\frac{bMC_{\rm total}}{\displaystyle \sum_{a=1}^{a=A} M_aC_{a,\rm detectable}}
\end{equation}

where $D$ is the total number of detections across all $p$ passes. Additionally, since the number of detections found can only be an integer number, the recorded value $D$ should actually be seen as a lower bound on the true number of detections $D_T$, where $D \leq D_T < D+1$. As such, the true number of objects, $N_T$, will lie in the range $N(D) \leq N_T < N(D+1)$.

\section{Results}
\label{sec:Results}


Based on the arguments above, it is possible to predict the number of RSOs in a particular region of orbital $h-i$ parameter space by observing a number of passes of a theoretical orbit. Depending on the latitude of the observing site and the tracked $h-i$ pair, the investigated region differs and the inferred population changes. To build up a full population of RSOs on all orbits would require significant investment of observing resources, tracking multiple passes of multiple theoretical orbits. To increase the resolution of the investigated $h-i$ regions, the observations are better taken from higher latitude observing sites. Table \ref{tab:results} gives the results of the above simulation for a range of $h/i/\delta$ combinations, as well as for a number of different maximum relative velocity thresholds and their corresponding magnitudes (using equation \ref{eq:M(v)}). The number of targets given assumes 1 detection per 20 passes, but can be simply scaled by the number of detections found (the chosen number of detections is simply an example, 20 passes was chosen as at least this number of fully observable passes per night is possible from most latitudes). The $h$, $i$, $\Omega$, $\nu$ region columns give the maximum positive and negative offsets from the tracked orbital parameters that produce a detectable target.

\begin{table*}
    \centering
    \caption{Table of predicted number of RSOs inferred from observations of a theoretical tracked orbit. First four columns give the tracked orbit parameters, observation latitude and limiting relative velocity used. Column five gives the corresponding limiting magnitude (using equation \ref{eq:M(v)}). Column six gives the the predicted number of RSOs assuming 1 detection in 20 passes. Remaining columns give maximum positive and negative orbital element offsets that produce detectable targets.}
    \label{tab:results}
    \begin{tabular}{ccccccccccccc}
    \hline
    $h$ (km) & $i$ ($\degree$) & $\delta$ ($\degree$) & max vel (pix/s) & magnitude & N targets & $h$ region (km) & $i$ region ($\degree$) & $\Omega$ region ($\degree$) & $\nu$ region ($\degree$) \\ \hline
    550 & 99 & 75 & 10.0 & 13.50 & 52510 -- 105020 & $h^{+48}_{-46}$ & $i^{+0.3}_{-0.3}$ & $\Omega^{+1.5}_{-1.5}$ & $\nu^{+0.5}_{-0.5}$ \\[5pt]
    750 & 99 & 50 & 2.5 & 16.24 & 405597 -- 811193 & $h^{+58}_{-54}$ & $i^{+0.9}_{-0.9}$ & $\Omega^{+1.2}_{-1.3}$ & $\nu^{+0.5}_{-0.5}$ \\[5pt]
    750 & 99 & 50 & 5.0 & 15.71 & 148832 -- 297665 & $h^{+58}_{-62}$ & $i^{+1.2}_{-1.2}$ & $\Omega^{+1.8}_{-1.8}$ & $\nu^{+0.6}_{-0.5}$ \\[5pt]
    750 & 99 & 50 & 7.5 & 14.59 & 82380 -- 164760 & $h^{+66}_{-64}$ & $i^{+1.2}_{-1.4}$ & $\Omega^{+1.9}_{-2.1}$ & $\nu^{+0.6}_{-0.6}$ \\[5pt]
    750 & 99 & 50 & 10.0 & 13.50 & 54370 -- 108740 & $h^{+72}_{-68}$ & $i^{+1.5}_{-1.6}$ & $\Omega^{+2.3}_{-2.4}$ & $\nu^{+0.7}_{-0.7}$ \\[5pt]
    750 & 99 & 75 & 10.0 & 13.50 & 20767 -- 41535 & $h^{+68}_{-66}$ & $i^{+0.4}_{-0.4}$ & $\Omega^{+1.7}_{-1.7}$ & $\nu^{+0.7}_{-0.7}$ \\[5pt]
    750 & 115 & 50 & 10.0 & 13.50 & 82984 -- 165969 & $h^{+82}_{-78}$ & $i^{+1.4}_{-1.6}$ & $\Omega^{+2.7}_{-2.9}$ & $\nu^{+1.0}_{-1.0}$ \\[5pt]
    850 & 95 & 75 & 10.0 & 13.50 & 31325 -- 62650 & $h^{+70}_{-68}$ & $i^{+0.4}_{-0.5}$ & $\Omega^{+1.5}_{-1.7}$ & $\nu^{+0.6}_{-0.6}$ \\[5pt]
    850 & 99 & 29 & 2.5 & 16.24 & 695484 -- 1390968 & $h^{+106}_{-98}$ & $i^{+2.2}_{-2.2}$ & $\Omega^{+2.1}_{-1.9}$ & $\nu^{+1.3}_{-1.1}$ \\[5pt]
    850 & 99 & 29 & 5.0 & 15.71 & 254061 -- 508121 & $h^{+138}_{-118}$ & $i^{+3.2}_{-3.0}$ & $\Omega^{+2.9}_{-2.5}$ & $\nu^{+1.6}_{-1.3}$ \\[5pt]
    850 & 99 & 29 & 7.5 & 14.59 & 131839 -- 263679 & $h^{+148}_{-126}$ & $i^{+3.6}_{-3.6}$ & $\Omega^{+3.0}_{-2.9}$ & $\nu^{+1.7}_{-1.4}$ \\[5pt]
    850 & 99 & 29 & 10.0 & 13.50 & 80392 -- 160784 & $h^{+156}_{-126}$ & $i^{+3.9}_{-4.0}$ & $\Omega^{+3.1}_{-3.2}$ & $\nu^{+1.7}_{-1.4}$ \\[5pt]
    850 & 99 & 50 & 2.5 & 16.24 & 287108 -- 574217 & $h^{+58}_{-58}$ & $i^{+0.9}_{-0.9}$ & $\Omega^{+1.3}_{-1.3}$ & $\nu^{+0.5}_{-0.4}$ \\[5pt]
    850 & 99 & 50 & 5.0 & 15.71 & 111222 -- 222444 & $h^{+62}_{-66}$ & $i^{+1.3}_{-1.3}$ & $\Omega^{+2.0}_{-2.0}$ & $\nu^{+0.6}_{-0.5}$ \\[5pt]
    850 & 99 & 50 & 7.5 & 14.59 & 60585 -- 121171 & $h^{+70}_{-70}$ & $i^{+1.5}_{-1.5}$ & $\Omega^{+2.3}_{-2.3}$ & $\nu^{+0.6}_{-0.6}$ \\[5pt]
    850 & 99 & 50 & 10.0 & 13.50 & 40188 -- 80377 & $h^{+78}_{-74}$ & $i^{+1.9}_{-1.8}$ & $\Omega^{+3.0}_{-2.8}$ & $\nu^{+0.7}_{-0.7}$ \\[5pt]
    850 & 99 & 75 & 10.0 & 13.50 & 16218 -- 32437 & $h^{+76}_{-72}$ & $i^{+0.4}_{-0.4}$ & $\Omega^{+1.7}_{-1.7}$ & $\nu^{+0.7}_{-0.7}$ \\[5pt]
    850 & 103 & 75 & 10.0 & 13.50 & 27063 -- 54127 & $h^{+78}_{-76}$ & $i^{+0.3}_{-0.3}$ & $\Omega^{+2.0}_{-2.0}$ & $\nu^{+0.9}_{-0.8}$ \\[5pt]
    950 & 99 & 75 & 10.0 & 13.50 & 13884 -- 27769 & $h^{+88}_{-86}$ & $i^{+0.4}_{-0.4}$ & $\Omega^{+2.0}_{-2.1}$ & $\nu^{+0.9}_{-0.8}$ \\[5pt]
    \hline
    \end{tabular}
\end{table*}


To correctly make inferences about the total number of targets in a particular orbital region we must first limit the relative velocity of any detections. The predicted number of targets is inferred from the number of detections but requires an assumption of an upper limit to the relative velocity of a target. From figures \ref{fig:injection_recovery_prob} and \ref{fig:mag_limits} however, we see that it is possible for a target moving faster than the imposed velocity threshold to be detected, if the target is sufficiently bright. Therefore, we must first filter our detections, rejecting any detection moving above the chosen relative velocity threshold. Since the relative velocity of a target changes over a single pass, this means we reject only those targets whose slowest detected relative velocity is above the threshold. This ensures the detections we count fall into the simulated parameter space, from which we can make inferences about the total number of targets.

Conversely, for targets moving below the relative velocity threshold, the limiting magnitude is not constant, and thus the magnitude to which the detection sample will be complete depends on the exact velocity for each target, not just the defined threshold. Thus, a target moving below the velocity threshold may or may not be detectable depending on its magnitude. To account for this, we must again filter our detection sample, rejecting detections for which $M>M(v)$, i.e. targets fainter than the magnitude corresponding to the defined relative velocity threshold. This ensures that our sample of detections is complete down to a magnitude of $M(v)$. The numbers given in column six of Table \ref{tab:results} are therefore interpreted as being the predicted number of targets within the given $h-i$ parameter space brighter than the given magnitude.

The result of this filtering approach means that we can filter our detections multiple times, each time producing a slightly different inferred population of targets. From a full set of detections from $p$ passes, we can select multiple different relative velocity thresholds, filter the detections accordingly, and thus infer multiple different RSO populations without requiring additional data. As such, Table \ref{tab:results} shows the results of using different velocity thresholds. As the chosen velocity threshold decreases, the magnitude to which the sample is complete becomes fainter, and the resolution on the $h-i$ region of parameter space becomes higher.

\section{Discussion and conclusions}
\label{sec:Discussion and conclusions}


The work presented here is a description and simulation of a novel method for inferring RSO populations by observing neighbouring orbits. By making certain assumptions, this technique allows one to infer the number of RSOs in particular $h-i$ orbital parameter bins. The method is statistical, extrapolating detections based on an underlying distribution, to determine the full population. A complete survey would require observing multiple passes while tracking orbits with a range of $h-i$ combinations. Ideally, this survey would also be carried out across multiple observing sites situated across a range of latitudes so as to analyse $h-i$ bins with a range of resolutions (as seen above, observing the same $h-i$ region at different latitudes allows inferences about different sized regions around the chosen $h-i$ pair).

Since the proposed method relies on detecting objects with low relative velocity, it can detect fainter than other methods at LEO. Objects in LEO have high orbital velocities and as such, spread their flux over a large number of pixels when observed with a stationary telescope. This can be improved upon using bespoke techniques (i.e. blind stacking, as discussed in \cite{COOKE2023907}) but the target streaks are still hundreds of pixels long in a single exposure. The above method detects neighbouring orbit targets with streaks less than 10 pixels per exposure. Depending on the chosen threshold for relative velocity, this method can achieve 50\% recovery of targets down to ${\sim}\,16^{\rm th}$ magnitude, a significant improvement over other methods tested on the same equipment and at the same observing site \citep{COOKE2023907}. The downside, of course, is that the parameter space of detections is much more constrained, only able to detect targets moving on similar orbits to the targeted one. Therefore, there is a trade-off between different methods, depending on the goals of the survey; a more complete population approach, or a targeted analysis of RSOs on specific orbits.

This manuscript lays out the neighbouring orbits technique in terms of simulations. The next step would be to begin a campaign of observations of a range of $h-i$ orbital pairs and make predictions based on detections. Once a set of detections has been established, the inferred number of detections can be calculated and compared to other estimates, generated through theoretical and/or simulated populations, resulting in an independent validation of these methods. A full survey of all potential orbital heights and inclinations would be a significant undertaking, but focusing on a smaller subset of $h-i$ pairs of particular interest would be more feasible. A future work for this project will be to begin a campaign of this type, most likely situated at the observing site on La Palma ($\delta=29\degree$). Collaboration with, or data from, other observing sites would allow for a more complete picture of the orbital distribution of RSOs to be obtained.

\section*{Acknowledgements}
\label{sec:Acknowledgements}

BFC and PC acknowledge support from a Science and Technology Facilities Council CLASP award (grant ST/V002279/1) and from the Defence Science and Technology Laboratory (UK). JAB gratefully acknowledges support from the Defence Science and Technology Laboratory (UK) and the Science and Technology Facilities Council (grant ST/Y50998X/1).

This work has made use of data obtained using the Warwick CLASP test telescope operated on the island of La Palma by the University of Warwick in the Spanish Observatory del Roque de los Muchachos of the Instituto de Astrofísica de Canarias.

The authors would like to thank an anonymous referee for their useful comments.

For the purpose of open access, the author has applied a Creative Commons Attribution (CC-BY) licence to any Author Accepted Manuscript version arising from this submission.

\section*{Data Availability}

Relevant data are freely available under a CC-BY-4.0 license at \href{https://github.com/BenCooke95/RSO\_populations}{https://github.com/BenCooke95/RSO\_populations}. Further details may be granted upon reasonable request to the corresponding author.



\bibliographystyle{rasti}
\bibliography{manuscript_new_clean} 




\appendix

%
\onecolumn
\section{Orbital parameter derivations}
\label{sec:Orbital parameter derivations}


\subsection{Definitions}

Let the relevant orbit be circular, with radius $r$ and inclination $\theta$. The orbit should pass through the zenith of the observing site at the chosen epoch.\\
Let the observing site, have $\rm{latitude}=\delta$, $\rm{longitude}=\alpha$.\\
Let a defined orbit cross the equator in the positive z-direction at $(r, \frac{\pi}{2}, \phi_e)$.\\
Let $\phi_{RAAN}$ be the Right Ascension of the Ascending Node (RAAN) of a polar orbit that passes directly over observing site.\\
Orbital parameter $\Omega$ is equal to the RAAN of the observing site (equal to the sidereal time) plus the difference between longitude of the observing site and longitude of the point at which the orbit crosses the equator in the positive z-direction. Orbital parameter $\nu$ is equal to the great circle angle between the observing site and the point at which the orbit crosses the equator in the positive z-direction.



\begin{figure}
    \centering
    \begin{subfigure}{0.49\textwidth}
        \centering
        \includegraphics[width=\textwidth]{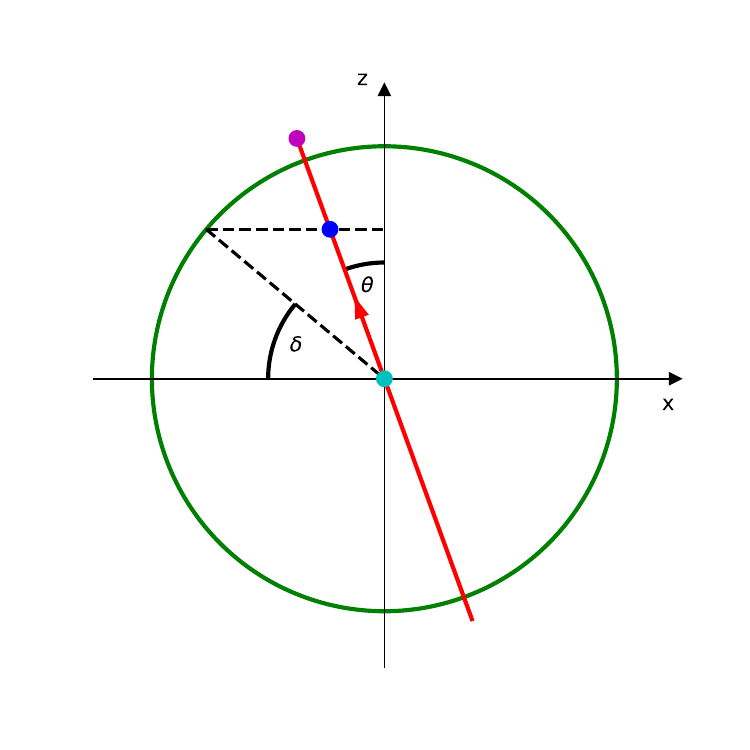}
        \caption{Orbital projection from $\theta = \frac{\pi}{2}$, $\phi = \phi_e$ (equator).}
        \label{fig:orbital_projection_1}
    \end{subfigure}
    \begin{subfigure}{0.49\textwidth}
        \centering
        \includegraphics[width=\textwidth]{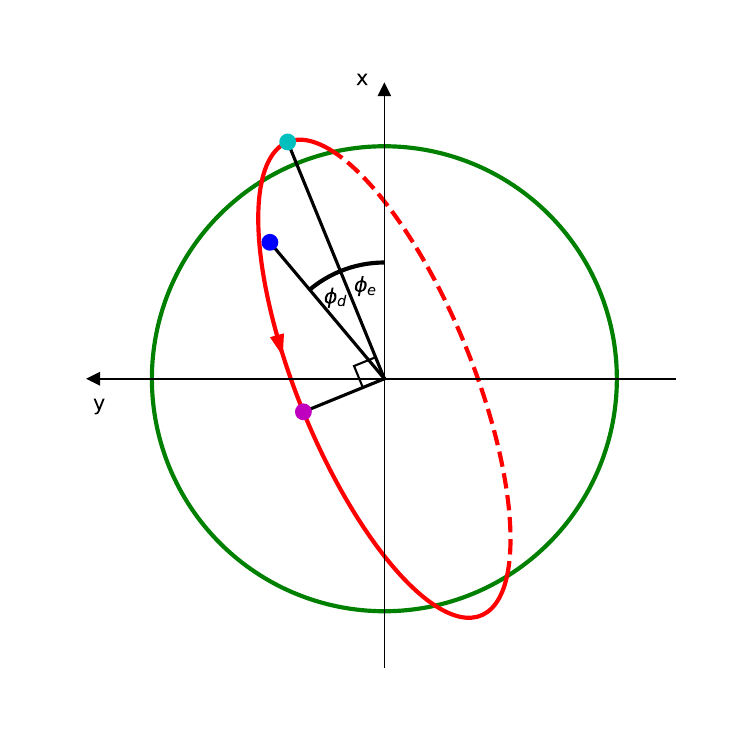}
        \caption{Orbital projection from $\theta = 0$, $\phi = 0$ (pole).}
        \label{fig:orbital_projection_2}
    \end{subfigure}
    \caption{Orbital projections. Green represents the Earth, black, the Cartesian coordinate axes and construction lines, and red, the orbit, with an arrow showing direction of motion (dotted portion shows orbit behind the earth). Three individual points are also shown; blue, for the observing site, magenta, for the orbital point with maximum z-coordinate, and cyan, for the point at which the orbit cross the equator in the positive z-direction. The observing site in Fig. \ref{fig:orbital_projection_2} appears to lie off the line of the orbit due to angular projection effects. Projections are constructed using $\theta=20^\degree$, $\alpha=\delta=40^\degree$, $r=1.1R_\oplus$ for illustration purposes.}
    \label{fig:orbital_projections}
\end{figure}

\subsection{Derivations}

Define a circular orbit as the intersection of a sphere of radius $r$ at the origin, and a plane.
To define a plane, we need three distinct points on plane:\\
\indent Origin: $P_1 = (x_1, y_1, z_1) = (0, 0, 0)$\\
\indent Observing site: $P_2 = (x_2, y_2, z_2) = (r, \theta_2, \phi_2)$\\
\indent Point in orbit at maximum z: $P_3 = (x_3, y_3, z_3) = (r, \theta_3, \phi_3)$

\noindent In Cartesian coordinates:

\begin{equation*}
\label{eq:1}
    P_1 = \begin{pmatrix} x_1 \\ y_1 \\ z_1 \end{pmatrix} = \begin{pmatrix} 0 \\ 0 \\ 0 \end{pmatrix}
\end{equation*}
\begin{equation*}
\label{eq:2}
    P_2 = \begin{pmatrix} x_2 \\ y_2 \\ z_2 \end{pmatrix} = \begin{pmatrix} r\cos{\phi_2}\sin{\theta_2} \\ r\sin{\phi_2}\sin{\theta_2} \\ r\cos{\theta_2} \end{pmatrix}
\end{equation*}
\begin{equation*}
\label{eq:3}
    P_3 = \begin{pmatrix} x_3 \\ y_3 \\ z_3 \end{pmatrix} =  \begin{pmatrix}r\cos{\phi_3}\sin{\theta_3} \\ r\sin{\phi_3}\sin{\theta_3} \\ r\cos{\theta_3} \end{pmatrix}
\end{equation*}

\noindent Equation of plane, $P$, from $P_1, P_2, P_3$:

\begin{equation}
\label{eq:4}
    P = (P_2-P_1) \times (P_3-P_1) = P_2 \times P_3 = r^2\begin{pmatrix} \sin{\phi_2}\sin{\theta_2}\cos{\theta_3} - \sin{\phi_3}\sin{\theta_3}\cos{\theta_2} \\ \cos{\phi_3}\sin{\theta_3}\cos{\theta_2} - \cos{\phi_2}\sin{\theta_2}\cos{\theta_3} \\ \cos{\phi_2}\sin{\theta_2}\sin{\phi_3}\sin{\theta_3} - \sin{\phi_2}\sin{\theta_2}\cos{\phi_3}\sin{\theta_3} \end{pmatrix}
\end{equation}

\noindent For any point $Q(x, y, z) = Q(r, \theta, \phi)$ on plane $P$:

\begin{equation}
\label{eq:5}
    P \cdot Q = 0.
\end{equation}


\subsubsection{Calculation of $\phi_e$ and $n$}

\noindent Let $Q$ be point at which orbit crosses equator in positive z direction. Thus $Q(x, y, z) = Q(r, \frac{\pi}{2}, \phi_e)$.

\begin{equation}
\label{eq:6}
    Q = \begin{pmatrix} x \\ y \\ z \end{pmatrix} = \begin{pmatrix} r\cos{\phi_e}\sin{\frac{\pi}{2}} \\ r\sin{\phi_e}\sin{\frac{\pi}{2}} \\ r\cos{\frac{\pi}{2}} \end{pmatrix} = \begin{pmatrix} r\cos{\phi_e} \\ r\sin{\phi_e} \\ 0 \end{pmatrix}
\end{equation}

\noindent Thus, substituting equations \ref{eq:4} and \ref{eq:6} into equation \ref{eq:5}:

\begin{equation}
\label{eq:7}
    r^2\begin{pmatrix} \sin{\phi_2}\sin{\theta_2}\cos{\theta_3} - \sin{\phi_3}\sin{\theta_3}\cos{\theta_2} \\ \cos{\phi_3}\sin{\theta_3}\cos{\theta_2} - \cos{\phi_2}\sin{\theta_2}\cos{\theta_3} \\ \cos{\phi_2}\sin{\theta_2}\sin{\phi_3}\sin{\theta_3} - \sin{\phi_2}\sin{\theta_2}\cos{\phi_3}\sin{\theta_3} \end{pmatrix} \cdot \begin{pmatrix} r\cos{\phi_e} \\ r\sin{\phi_e} \\ 0 \end{pmatrix} = 0
\end{equation}

\begin{equation}
\label{eq:8}
    r^3[\sin{\phi_2}\sin{\theta_2}\cos{\theta_3}\cos{\phi_e} - \sin{\phi_3}\sin{\theta_3}\cos{\theta_2}\cos{\phi_e} + \cos{\phi_3}\sin{\theta_3}\cos{\theta_2}\sin{\phi_e} - \cos{\phi_2}\sin{\theta_2}\cos{\theta_3}\sin{\phi_e}] + 0 = 0
\end{equation}

\noindent Note that $\theta_3 = \theta$. From symmetry, $\phi_3$ is exactly $\frac{\pi}{2}$ around equator from $\phi_e$:

\begin{equation*}
\label{eq:9}
    \phi_3 = \phi_e + \frac{\pi}{2}\ \ \therefore\ \sin{\phi_3} = \cos{\phi_e},\ \cos{\phi_3} = -\sin{\phi_e}
\end{equation*}

\noindent Thus:

\begin{equation*}
\label{eq:10}
    \sin{\phi_2}\sin{\theta_2}\cos{\theta}\cos{\phi_e} - \cos{\phi_e}\sin{\theta}\cos{\theta_2}\cos{\phi_e} - \sin{\phi_e}\sin{\theta}\cos{\theta_2}\sin{\phi_e} - \cos{\phi_2}\sin{\theta_2}\cos{\theta}\sin{\phi_e} = 0
\end{equation*}

\begin{equation*}
\label{eq:11}
    \sin{\phi_2}\sin{\theta_2}\cos{\theta}\cos{\phi_e} = \cos{\phi_2}\sin{\theta_2}\cos{\theta}\sin{\phi_e} + \sin{\theta}\cos{\theta_2}(\sin^2{\phi_e}+\cos^2{\phi_e})
\end{equation*}

\begin{equation}
\label{eq:12}
    \sin{\phi_2}\sin{\theta_2}\cos{\theta}\cos{\phi_e} = \cos{\phi_2}\sin{\theta_2}\cos{\theta}\sin{\phi_e} + \sin{\theta}\cos{\theta_2}
\end{equation}

\noindent Note that $P_2$ represents the observing site, thus $\phi_2=\alpha$, $\theta_2=\frac{\pi}{2} - \delta$. Therefore $\sin{\theta_2} = \cos{\delta}$, $\cos{\theta_2} = \sin{\delta}$. Substituting into equation \ref{eq:12} gives:

\begin{equation*}
\label{eq:13}
    \sin{\alpha}\cos{\delta}\cos{\theta}\cos{\phi_e} = \cos{\alpha}\cos{\delta}\cos{\theta}\sin{\phi_e} + \sin{\theta}\sin{\delta}
\end{equation*}

\begin{equation}
\label{eq:16}
    A\cos{\phi_e} = B\sin{\phi_e} + C
\end{equation}

\noindent Where $A = \sin{\alpha}\cos{\delta}\cos{\theta}$, $B = \cos{\alpha}\cos{\delta}\cos{\theta}$, $C = \sin{\theta}\sin{\delta}$, defined in terms of $\alpha$, $\delta$ and $\theta$.\\




\noindent Using the double angle formula:

\begin{equation}
\label{eq:17}
    A(\cos^2{\frac{\phi_e}{2}} - \sin^2{\frac{\phi_e}{2}}) = B(2\sin{\frac{\phi_e}{2}}\cos{\frac{\phi_e}{2}}) + C
\end{equation}

\noindent Let $\phi_{e2} = \frac{\phi_e}{2}$, therefore:

\begin{equation*}
\label{eq:19}
    A\cos^2{\phi_{e2}} - A\sin^2{\phi_{e2}} = 2B\sin{\phi_{e2}}\cos{\phi_{e2}} + C(\sin^2{\phi_{e2}} + \cos^2{\phi_{e2}})
\end{equation*}

\begin{equation*}
\label{eq:20}
    A - A\tan^2{\phi_{e2}} = 2B\tan{\phi_{e2}} + C\tan^2{\phi_{e2}} + C
\end{equation*}

\begin{equation*}
\label{eq:21}
    (C+A)\tan^2{\phi_{e2}} + 2B\tan{\phi_{e2}} + (C-A) = 0
\end{equation*}

\begin{equation*}
\label{eq:22}
    \tan{\phi_{e2}} = \frac{-2B\pm\sqrt{(2B)^2 - 4(C+A)(C-A)}}{2(C+A)}
\end{equation*}

\begin{equation}
\label{eq:23}
    \tan{\phi_{e2}} = \frac{\pm\sqrt{A^2 + B^2 - C^2} - B}{A+C}
\end{equation}

\noindent Thus, taking the positive root:

\begin{equation}
\label{eq:24}
    \phi_e = 2\tan^{-1}\left({\frac{\sqrt{A^2 + B^2 - C^2} - B}{A+C}}\right)
\end{equation}

\noindent Let $\phi_d = \alpha - \phi_e$. Thus, finally:

\begin{equation}
\label{eq:25}
    \Omega = \phi_{RAAN} + \phi_d \pmod{2\pi}.
\end{equation}

\subsubsection{Calculation of $\sigma$ and $\nu$}

\noindent Utilise the great circle angle formula (for example derivation see \cite{GreatCircleAngle}). $\theta$ measured from equator to pole, $\phi$ measured around equator. $\sigma$ is angle between $P(r, \theta_1, \phi_1)$ and $Q(r, \theta_2, \phi_2)$.

\begin{equation}
\label{eq:26}
    \sigma = \cos^{-1}\left(\sin{\theta_1}\sin{\theta_2} + \cos{\theta_1}\cos{\theta_2}\cos(|{\phi_2-\phi_1}|)\right)
\end{equation}

\noindent $P$ is point at which orbit crosses equator, thus $\theta_1 = 0$, $\phi_1 = \phi_e$. $Q$ is observing site, thus $\theta_2 = \delta$, $\phi_2 = \alpha$. Therefore:

\begin{equation*}
\label{eq:27}
    \sigma = \cos^{-1}\left(\sin{\delta}\sin{0} + \cos{\delta}\cos{0}\cos|{\alpha-\phi_e}|\right)
\end{equation*}

\begin{equation}
\label{eq:28}
    \sigma = \cos^{-1}\left(\cos{\delta}\cos|{\alpha-\phi_e}|\right)
\end{equation}

\begin{equation}
\label{eq:29}
    \nu = \sigma \pmod{2\pi}.
\end{equation}

\section{Target detectability comparison plots}
\label{sec:Target detectability comparison plots}

\begin{figure}
    \centering
    \begin{subfigure}{\columnwidth}
        \centering
        \includegraphics[width=0.5\columnwidth]{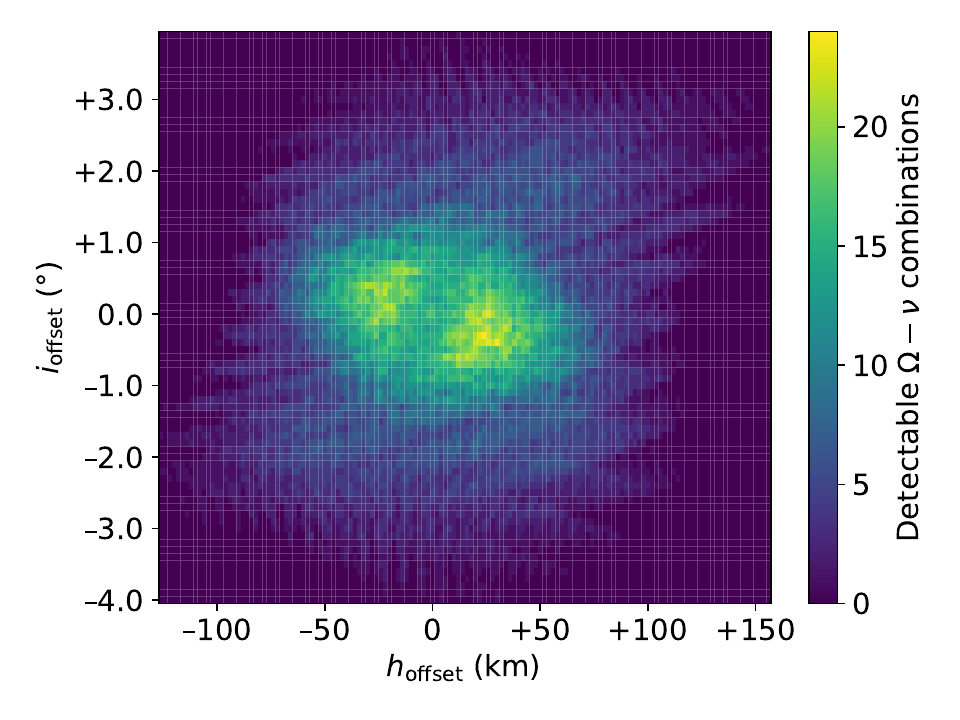}
        \caption{$\delta = 29\degree$}
        \label{fig:hi_detections_850_99_29_2}
    \end{subfigure}
    \begin{subfigure}{\columnwidth}
        \centering
        \includegraphics[width=0.5\columnwidth]{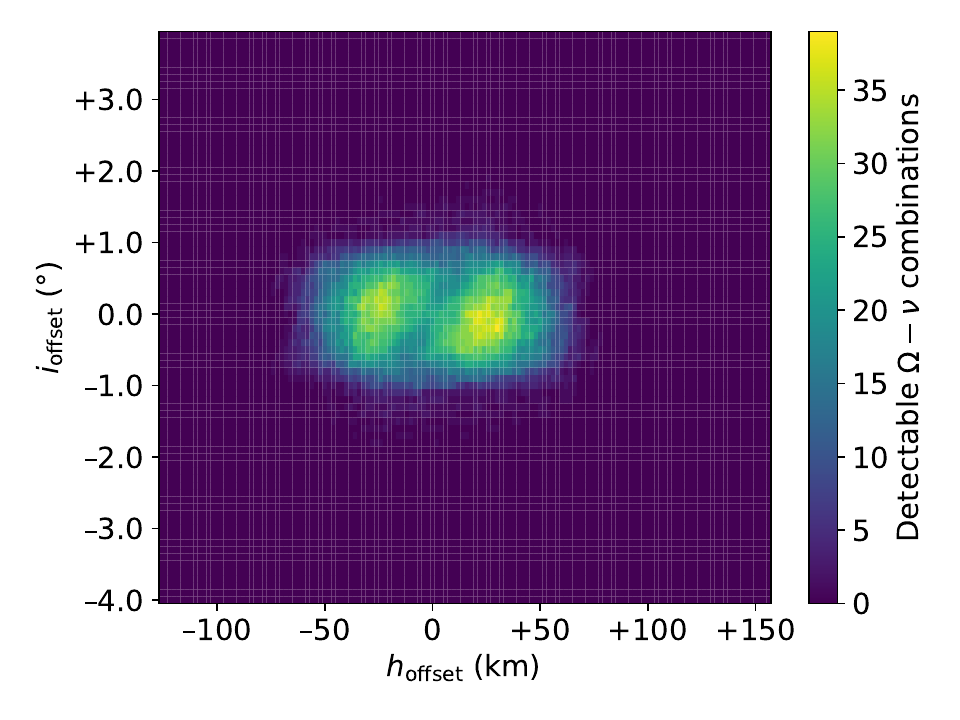}
        \caption{$\delta = 50\degree$}
        \label{fig:hi_detections_850_99_50_2}
    \end{subfigure}
    \begin{subfigure}{\columnwidth}
        \centering
        \includegraphics[width=0.5\columnwidth]{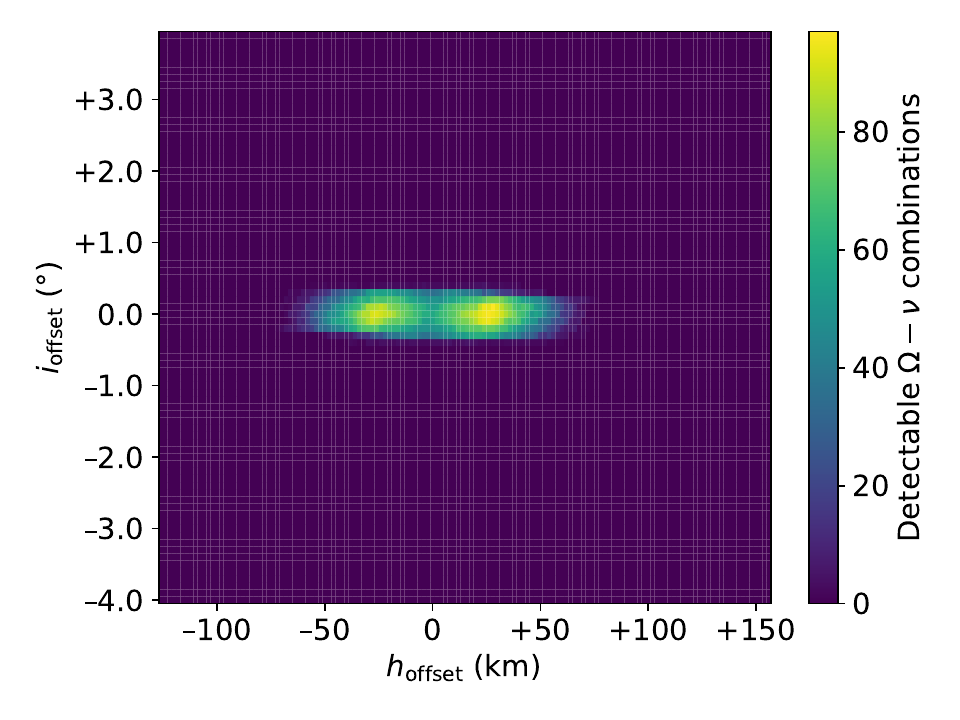}
        \caption{$\delta = 75\degree$}
        \label{fig:hi_detections_850_99_75_2}
    \end{subfigure}
    \caption{Reproduction of figures \ref{fig:hi_detections_850_99_29}, \ref{fig:hi_detections_850_99_50} and \ref{fig:hi_detections_850_99_75} on the same axis scale. Target detectability as a function of $h_{\rm offset}$ and $i_{\rm offset}$ for 
    $O_{\rm principal} = (850, 99, 29.8, 28.9)$. Colour designates relative probability, i.e. number of $\Omega-\nu$ combinations resulting in a detection for each $h-i$ pair. Values assume a limiting relative velocity of 10\,pix/s.}
    \label{fig:hi_detections_850_99_50/75_2}
\end{figure}


\bsp	
\label{lastpage}
\end{document}